\documentclass[%
 reprint,
 amsmath,amssymb,
 aps,
]{revtex4-2}
\usepackage[T1]{fontenc}
\usepackage{graphicx}
\usepackage{comment}
\usepackage{xcolor}
\usepackage{dcolumn}
\usepackage{bm}
\usepackage{hyperref}%

\usepackage{amsmath} 
\usepackage{amssymb}
\usepackage{amsfonts}
\usepackage{mathrsfs}
\usepackage{braket}

\newcommand{\db}{2\nu\beta\beta}
\newcommand{\nm}{\mathcal{M}_{\db}}
\newcommand{\ca}{{}^{48}\text{Ca}}
\newcommand{\scn}{{}^{48}\text{Sc}}
\newcommand{\ti}{{}^{48}\text{Ti}}

\newcommand{\iso}{\tau^{\pm}}

\newcommand{\psii}{\psi_{i}}
\newcommand{\psif}{\psi_{f}}

\newcommand{\htapm}{\hat{\tau}^{\pm}}
\newcommand{\spn}{\hat{\sigma}}
\newcommand{\du}{\gamma^{\mu}}
\newcommand{\dd}{\gamma_{\mu}}
\newcommand{\vud}{V_{\text{ud}}}

\newcommand{\be}{\begin{equation}}
\newcommand{\ee}{\end{equation}}
\newcommand{\ba}{\begin{array}}
\newcommand{\ea}{\end{array}}
\newcommand{\bn}{\begin{eqnarray}}
\newcommand{\en}{\end{eqnarray}}

\begin{document}

\preprint{APS/123-QED}
\title{Two-neutrino $0^+ \rightarrow 0^+$ double beta decay of ${}^{48}\text{Ca}$ within the DFT-NCCI framework}

\author{Jan Miśkiewicz}
\affiliation{Faculty of Physics, University of Warsaw, Poland}

 \author{Maciej Konieczka}
\affiliation{Faculty of Physics, University of Warsaw, Poland}
 
 \author{Wojciech Satuła}
\affiliation{Faculty of Physics, University of Warsaw, Poland}

\date{\today}

\begin{abstract}
We present a seminal calculation of the nuclear matrix element for the two-neutrino double beta ($2\nu\beta\beta$) decay of ${}^{48}\text{Ca} \rightarrow {}^{48}\text{Ti}$ using a post-Hartree-Fock (HF) Density Functional Theory-based No-Core Configuration-Interaction (DFT-NCCI) framework developed by our group [Phys. Rev. C 94, 024306 (2016)]. In the present calculation, we utilize a variant of the approach that restores rotational symmetry and mixes states projected from self-consistent mean-field configurations obtained by solving the HF equations with the density-independent local Skyrme interaction.
Our calculations yield $|\mathcal{M}_{2\nu\beta\beta}| = 0.056(6)$ MeV$^{-1}$ for the nuclear matrix element describing this process. This result is in very good agreement with shell-model studies - for example, with the calculations by Horoi {\it et al.\/} [Phys. Rev. C 75, 034303 (2007)], which yielded 0.054 (0.064) MeV$^{-1}$ for the GXPF1A (GXPF1) interactions, respectively. It is also in a reasonable agreement with the most recent experimental estimate from the review by Barabash, which is 0.068(6) MeV$^{-1}$, assuming quenching $qg_\text{A} \approx 1$.
The consistency of our prediction with the shell-model results increases our confidence in the nuclear modeling of this second-order, very rare process which is of paramount importance for further modeling of the neutrinoless double beta ($0\nu\beta\beta$) decay process. 
\end{abstract}

\maketitle


\section{\label{sec:level1}Introduction}
Two-neutrino double beta decay ($\db$) is a second-order weak-interaction process. Consequently, it 
is among the rarest radioactive processes observed in nature. The theoretical relationship between the 
measured half-life $T_{1/2}$ and the quantum mechanical probability of the decay is given by \cite{(Avi08),(Suh98)}:
\begin{equation}
\label{eq1}
    T_{1/2}^{-1} = G_{2\nu\beta\beta} \cdot |\nm|^2,
\end{equation}
where $G_{2\nu\beta\beta}$ denotes the kinematic leptonic phase-space factor, and $\nm$ is the nuclear 
matrix element containing cumulative nuclear-structure-dependent information about transition rates between 
nuclear quantum states active in the process. \\

The $\db$ decay, although well known, has recently attracted significant attention due to substantial investments 
in the search for the yet unobserved neutrinoless double beta decay ($0\nu\beta\beta$), a process considered 
a potential gateway to {\it new physics} beyond the Standard Model. Current efforts focus on high-precision half-life 
measurements and, on the theoretical side, on high-precision calculations of the nuclear matrix elements using 
various theoretical models. 

The diversity of theoretical approaches to modeling $\db$ decay is of paramount importance. Agreement among predictions 
from different models would naturally strengthen our confidence that the nuclear modeling of this second-order, extremely 
rare process is well understood. Conversely, discrepancies between predictions would raise important questions about potentially 
missing physics. Both aspects are crucial for the continued development of reliable models for $0\nu\beta\beta$ decay.
 \\

This article is organized as follows. In Sect.~\ref{bb-decay}, we briefly overview the $\db$ decay. 
Sect.~\ref{DFT-NCCI} provides a concise description of the DFT-rooted No-Core Configuration-Interaction (DFT-NCCI) framework 
used to compute the $\db$ nuclear matrix element (NME), with particular attention given to the technical aspects 
of calculating the Gamow-Teller matrix element within the variant of the DFT-NCCI approach that involves only angular-momentum 
projection. In Sect.~\ref{config}, we describe in detail the configuration spaces for the $^{48}$Ca, $^{48}$Sc, and $^{48}$Ti nuclei. 
Sect.~\ref{results} presents the results of the numerical calculations. Finally, the summary and concluding remarks are 
provided in Sect.~\ref{summary}.

\section{Brief overview of $\db$-decay}\label{bb-decay}

The goal of this section is to provide, for the sake of self-consistency, a concise description of $\db$ decay necessary to understand the physics of this phenomenon. The readers interested in further details are referred, for example, to the dedicated reviews in Refs.\cite{(Doi85),(Tom91)} or to the textbook by Konopiński~\cite{(Kon66)}, which contains probably the most detailed derivation of the $\db$-decay process.

Beta decay is governed by the weak interaction Hamiltonian $\hat{H}_{\beta}$, whose density is given by~\cite{(Sev06)}:
\begin{equation}
    \label{hw1}
    \hat{\mathcal{H}}_{\beta} = \frac{G_F V_{\text{ud}}}{\sqrt{2}}\iso J^{\mu\dag} j_{\mu} + \text{h.c.}\ ,
\end{equation}
where $\iso$ is the isospin ladder operator, $J^\mu$ ($j_{\mu}$) denotes the quark (leptonic) four-current, $V_{\text{ud}}$ is the relevant element of the Cabibbo-Kobayashi-Maskawa matrix, and $G_F/(\hbar c)^3 = 1.1663788(6) \times 10^{-5}\ \text{GeV}^{-2}$ is the Fermi coupling constant~\cite{(Erl22)}. The precise definition of $G_F$ reflects the underlying mechanism of $\beta$ decay, which involves the exchange of massive $\text{W}^{\pm}$ bosons and is characterized by the weak-interaction coupling strength $g_W$:
\begin{equation}
    \label{eq2}
    \frac{G_F}{\sqrt{2}} = \frac{g_W^2}{8(m_W c^2)^2}.
\end{equation}
In the so-called impulse approximation, the $\beta$ decay of an $(A,Z)$ nucleus is assumed to occur on a single weakly interacting nucleon, with no influence on the remaining $A-1$ spectator nucleons~\cite{(Suh06)}. The small momentum transfer involved suppresses nucleonic excitations, allowing the replacement of quark spinors by their hadronic equivalents. Consequently, the effective hadronic four-current:
\begin{equation}
    \label{Jhad}
    J^{\mu \dag} = \Bar{\text{u}}\du(1-\gamma_5)\text{d},
\end{equation}
can be written in its nucleonic form as
\begin{equation}
        \label{Jnuc}
        J^{\mu \dag} = \Bar{\text{p}}\htapm\du(1-\gamma_5)\text{n}.
    \end{equation}
Multiplying nucleonic and leptonic currents and constraining products to the vector- and axial-vector-type, the density of the $\beta$ interaction may be expressed as:
\begin{widetext}
\begin{equation}
    \label{h_rel}
    \hat{\mathcal{H}}_{\beta} = \frac{G_F\cdot\vud}{\sqrt{2}}\big[ g_V \Bar{\text{e}}\du(1-\gamma_5)\nu_{e} \Bar{\text{p}}\htapm\dd\text{n} + g_A \Bar{\text{e}}\du(1-\gamma_5)\nu_{e} \Bar{\text{p}}\htapm\dd\gamma_5\text{n} \big] + \text{h.c.}\ ,
\end{equation}
\end{widetext}
where according to the CVC hypothesis, the hadronic form-factors $g_V=1$ and $g_A \approx -1.27$ \cite{(Suh06), (Sev06), (Her01)}.

Due to the relatively small magnitude of the weak force compared to the strong interaction that defines nuclear eigenstates, it can be treated as a perturbation within the framework of Fermi’s golden rule: 
\begin{equation}
    \label{eq3}
    d\lambda 
    = \frac{2\pi}{\hbar}| \langle \psif \vert \hat{H}_{\beta} \vert \psii \rangle |^2 
\end{equation}
Assuming that the leptonic particles are emitted as $s$-waves and that the nucleus can be described within the impulse approximation, the decay rate can be expressed as:
\begin{eqnarray}
    \lambda = \frac{G_F^2 f(Z,Q_{\beta})}{2\pi^3 c^6\hbar^7}\cdot |\mathcal{M}_{\beta}|^2,
\label{eq4}
\end{eqnarray}
where $f(Z,Q_{\beta})$ is the Fermi form factor, $Q_{\beta}$ is the $\beta$-decay $Q$-value, and  $\mathcal{M}_{\beta}$ is the reduced nuclear matrix element for the $\beta$-transition.

In the case of $\db$-active $(Z,A)$ nuclei, a single $\beta$ transition to $(Z+1,A)$ is forbidden. 
However, the nucleus can undergo a second-order process, decaying directly to the $(Z+2,A)$ nucleus.
\begin{figure}[h]
\includegraphics[scale = 0.50]{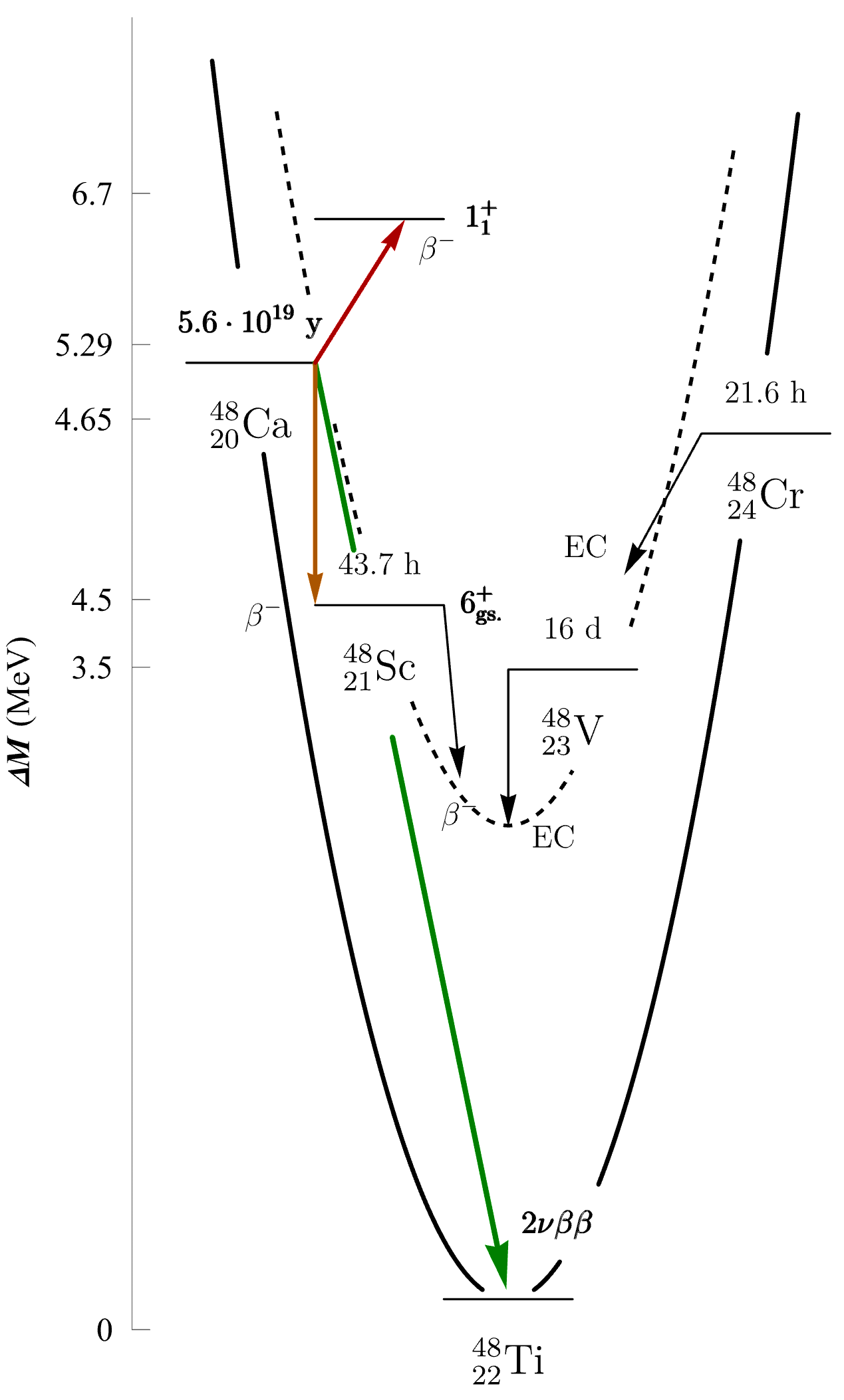}
\caption{\label{parabolas} 
Mass parabolas of even-even and odd-odd nuclei with mass number $A = 48$, showing allowed transitions and the corresponding half-lives. 
For the isotope $^{48}_{20}\text{Ca}$, single $\beta$ decay to  $^{48}_{21}\text{Sc}$ (orange arrow) is energetically allowed, but highly forbidden 
due to a large $\Delta I$ difference. According to the calculations in Ref.~\cite{(Aun99)}, the theoretical
$\beta$-decay half-life of $^{48}$Ca is estimated to be approximately 25 times longer than the measured double-beta-decay half-life (green arrow), whose extreme rarity is reflected in its very long half-life of 5.9 10$^{19}$ years. Image concept adapted from\cite{(Vog92)}; half-lives taken from~\cite{(Kon21a)}.} 
\end{figure}
The decay rate of $\db$ can also be estimated using Fermi’s golden rule. However, since single-$\beta$ transitions are forbidden, the process must be treated as a second-order perturbation:
\begin{eqnarray}
    \label{eq6}
    d\lambda = \frac{2\pi}{\hbar} \bigg|& & \sum_{m}\frac{\langle \psi_{f} \vert \hat{H}_{\beta} \vert \psi_{m} \rangle \langle \psi_{m} \vert \hat{H}^{\beta} \vert \psi_{i} \rangle }{E_i - E_m} \bigg|^2 \\
    \cdot& & \delta(E_f - E_i)d\mathcal{E}d\theta,    
\end{eqnarray}
where the measure $d\mathcal{E}$ refers to the differential density of leptonic states in the relativistic treatment:
\begin{equation}
    \label{eq00}
d\mathcal{E}=c^2 p_{1}p_{2}\varepsilon_{1}\varepsilon_{2}d\epsilon_{1}d\epsilon_{2}\omega_1^2\omega_2^2 d\omega_{1}d\omega_{2}.
\end{equation}
In (\ref{eq6}) $d\theta$ denotes the angle between the emitted leptonic pair, $Ei$ ($E_f$) are the initial (final) energies of the system, $p_i (\varepsilon_i$) are the momenta (energies) of the emitted electrons, and $\omega_i$ are the energies of the  neutrinos.

The expression in Eq.(\ref{eq6}) must be antisymmetrized with respect to the electron–antineutrino pairs and factorized into purely leptonic and nuclear components. Integration over the leptonic variables and the emission angle $\theta$ yields the $\db$ decay rate formula~\cite{(Suh98),(Kon66)}:
\begin{eqnarray}
    \label{dr2v}
      \lambda =  \frac{g^4_{A}(G_F V_{\text{ud}})^4}{96\pi^7\hbar^{13} m_e^2}|\nm|^2 && \int f(Z_f,\varepsilon_1) f(Z_f,\varepsilon_2) \cdot \nonumber \\ &&  \delta(E_f - E_i)d\mathcal{E}.
\end{eqnarray}
where $f(Z_f,\varepsilon_i)$  are the Fermi functions accounting for the Coulomb distortion of the outgoing electrons, and 
$\nm$ is the nuclear matrix element for the two-neutrino double beta decay. Due to the strong suppression of Fermi transitions 
in $\db$ decay, only the axial-vector component of the weak Hamiltonian $\hat{H}_{\beta}$ contributes significantly to the matrix element in Eq.~(\ref{eq6}).

The separation of leptonic and nuclear variables, achieved by antisymmetrizing the leptonic pairs and adopting the non-relativistic limit for the Gamow-Teller (GT) operators, results in the half-life formula in Eq.~(\ref{eq1}), with the leptonic phase space factor:
\begin{eqnarray}
    \label{psf}
      G_{\db}(Z,E)= \frac{g^4_{A}(G_F V_{\text{ud}})^4}{96\pi^7\hbar^{13} m_e^2} \int && f(Z_f,\varepsilon_1)f(Z_f,\varepsilon_2)\cdot \nonumber \\ && \delta(E_f - E_i)d\mathcal{E}
\end{eqnarray}
and the nuclear matrix element for $\db$ ($0^+ \rightarrow 0^+$) is given by:
\begin{equation}
    \label{eq9}
    \nm = \sum_{m} \frac{\langle 0^{+}_{f} \vert \sum_{i}\spn_{i}\htapm_{i} \vert 1_{m}^{+} \rangle \langle 1_{m}^{+} \vert \sum_{i}\spn_{i}\htapm_{i} \vert 0^{+}_{i} \rangle}{\Delta E_{m} + \frac{1}{2}Q_{\beta\beta}+\Delta M},
\end{equation}
where the sum runs over the $m$-intermediate states. The transition operators are GT operators: the $\pm$ sign corresponds to $\beta^{-}$/(${\beta^{+}}$, EC) decays, $Q_{\beta\beta}$ is the $\db$ $Q$-value, $\Delta E_m$ is the excitation energy of the $m$th intermediate 
state relative to the nucleus' ground state, and $\Delta M = M_{\text{int}} - M_i$ is the mass difference between the intermediate ($M_{\text{int}}$)
and mother ($M_i$) nuclei. 
\begin{figure}[h]
\includegraphics[scale = 0.65]{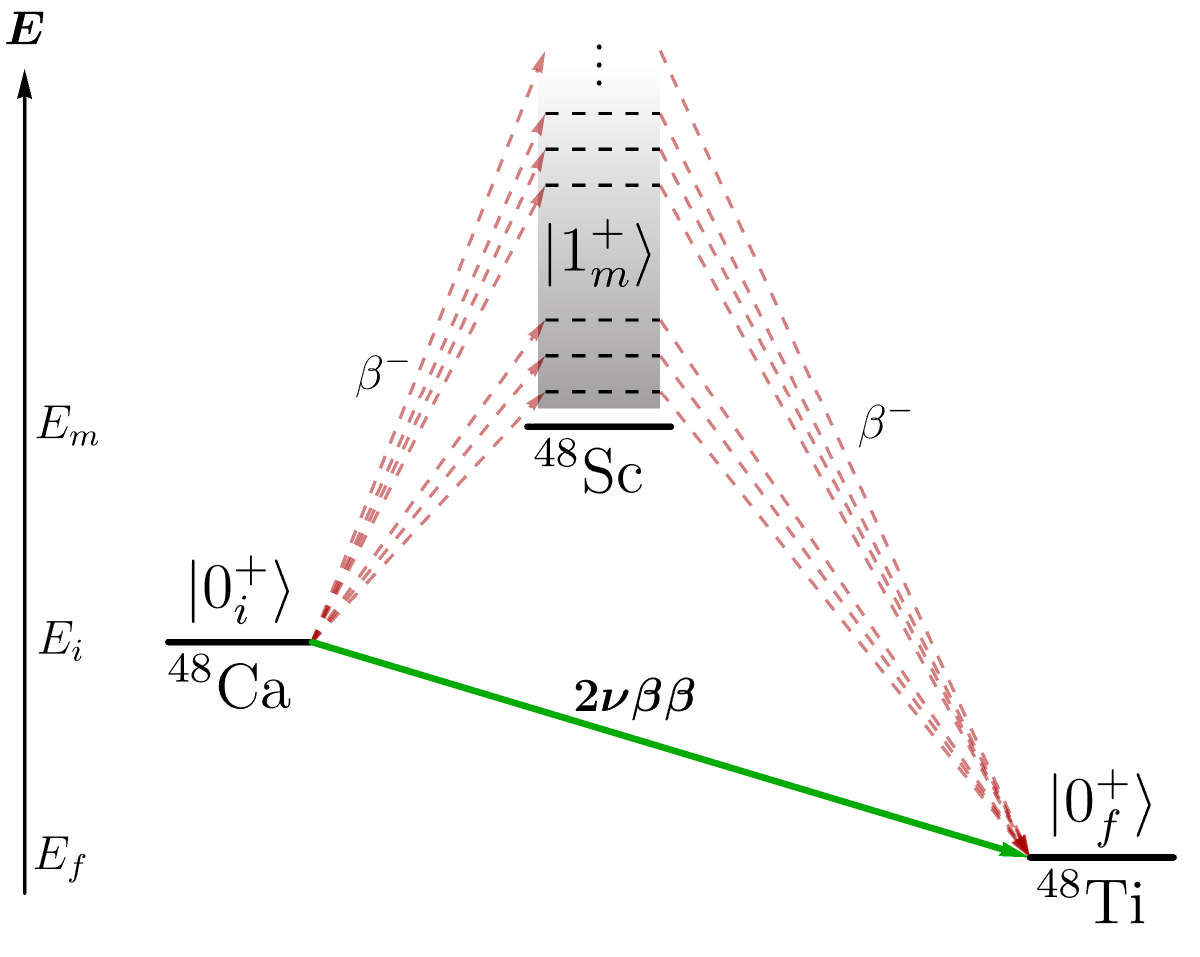}
\caption{\label{fermi}  
Schematic representation of the $\db$ decay of $\ca$. Although the single $\beta$ transitions (dashed red arrows) are energetically forbidden  they contribute virtually to the total decay rate.}
\end{figure}
The exact formula is valid only if the intermediate states form a complete basis set. However, in practice, the summation is performed until the matrix element saturates, i.e., when including higher-energy contributions no longer changes the total matrix element within an assumed uncertainty.

The evaluation of the $G_{\db}$ phase space factor has been the subject of parallel discussions due to its strong dependence on the theoretical treatment of the interaction between the nucleus and the emitted leptons. This problem is far from trivial and has been investigated in numerous studies, including~\cite{(Suh98), (Kot12), (Sto13x)}.

\section{DFT-NCCI framework}\label{DFT-NCCI}

The DFT-NCCI approach used here is a beyond-mean-field framework originally designed with the aim of studying 
isospin-symmetry breaking in $N\sim Z$ nuclei~\cite{(Sat16d)}. The approach restores rotational symmetry, treats 
isospin rigorously, and mixes states projected from self-consistent mean-field configurations obtained by solving the HF 
equations with a density-independent Skyrme pseudo-potential. In this work, we use the SV$_{\rm T}$ pseudo-potential, 
which is the SV density-independent Skyrme force of Ref.~\cite{(Bei75)} augmented with tensor terms. The use of a 
pseudo-potential is indispensable in avoiding the problem of singularities in the calculation of the Hamiltonian 
kernels~\cite{(Dob07)}.

The model has proven to be very successful in reproducing and predicting a variety of (pseudo-)observables, ranging from isospin impurities~\cite{(Sat09)} and isospin-symmetry-breaking (ISB) corrections to superallowed Fermi $\beta$-decays~\cite{(Sat11),(Kon22)}, 
through mirror- and triplet-energy differences~\cite{(Bac18),(Bac19)} in binding energies, to mirror-energy differences 
(MED) in $T=1/2$ $fp$-shell mirror nuclei~\cite{(Bac21)}, $T=1/2$ $^{79}$Zr/$^{78}$Y mirrors~\cite{(Lle20)}, and $T=3/2$ $A=45$ and $A=47$ mirrors~\cite{(Uth22),(Sat23)}. It has also proven to be a very effective tool for studying various aspects of Gamow-Teller 
$\beta$-decay, including the GT strength distributions in ${}^8\text{Li}$, ${}^8\text{Be}$, and ${}^{24}\text{Mg}$, see Ref.~\cite{(Kon18)}, 
or the GT matrix elements in $N\sim Z$ $sd$- and $pf$-shell nuclei where the model's predictions appeared to be comparable to the state-of-the-art nuclear shell models, see~\cite{(Kon16)}. 

The aim of this work is to demonstrate that the approach is capable of accounting for the very rare $2\nu\beta\beta$ decay of ${}^{48}\text{Ca} \rightarrow {}^{48}\text{Ti}$, which, being a second-order perturbative process, proceeds between the $|0^+\rangle$ ground states of the mother and daughter nuclei through the virtual intermediate $|1^+\rangle$ states in $\scn$, as schematically depicted in Fig.~\ref{fermi}.

Most of the applications of the DFT-NCCI model have concerned $N \sim Z$ nuclei and, therefore, were done with the full version of the model, including both the restoration of rotational symmetry and a rigorous treatment of isospin. In the case of the $2\nu\beta\beta$ decay of ${}^{48}\text{Ca}$, however, the isospin of the mother nucleus is $T = 8$, which justifies the use of a simplified variant of the model involving only angular-momentum projection. In this variant of our model, to compute the wave functions in all participating nuclei, we proceed as follows:
\begin{enumerate}
\item
In the first step, we solve self-consistently HF equations in order to compute a set of 
$N_{\rm conf}$ relevant low-lying (multi)particle-(multi)hole  
configurations $\{ \ket{\varphi_j}\}_{j=1}^{N_{\rm conf}}$. These configurations span 
the so called {\it configuration space\/}. 
\item
In the next step, we apply the  angular-momentum  
projection: 
\begin{eqnarray}
\hat P^I_{M K} & = & \frac{2I+1}{8\pi^2 } \int d\Omega\;
\; D^{I\, *}_{M K}(\Omega )\; e^{-i\gamma \hat{J}_z}
e^{-i\beta \hat{J}_y} e^{-i\alpha \hat{J}_z} ,
\end{eqnarray}
to each configuration $\hat{P}^{I}_{MK}\ket{\varphi_j}$  and perform
$K$-mixing, where $K$ stands for a projection of angular momentum onto the intrinsic $z$-axis. 
This gives us a set of good-angular-momentum non-orthogonal states 
\begin{equation}\label{m-space}
\ket{\varphi_j ; I M; T_z}^{(i)}=\frac{1}{\sqrt{\mathcal{N}^{(i)}_{\varphi_j ; IM;T_z}}}
\sum_{\substack{K }} a_{K}^{(i)} \hat{P}^{I}_{MK}\ket{\varphi_j}.
\end{equation} 
These states form the so called  {\it model space\/}.
\item
In the last step, we perform a mixing of the $\{\ket{\varphi_j ; I M; T_z}^{(i)}\}$ states 
for all configurations $\{ \ket{\varphi_j}\}_{j=1}^{N_{\rm conf}}$. Since the $\{\ket{\varphi_j ; I M; T_z}^{(i)}\}$ states 
are non-orthogonal, the mixing requires solving the Hill-Wheeler-Griffin (HWG) equation. 
In effect, we obtain a set of linearly 
independent  DFT-NCCI eigenstates of the form:
\begin{equation}
\ket{\psi_{\textrm{NCCI}}^{k; IM; T_z}}
=\frac{1}{\sqrt{\mathcal{N}^{(k)}_{IM;T_z}}}
\sum_{ij}c_{ij}^{(k)}\ket{\varphi_j;I M;T_z}^{(i)} \,. \label{eq:nccistate}
\end{equation}  
\end{enumerate}

The overcompleteness of the set of states ${\ket{\varphi_j ; I M; T_z}^{(i)}}$ implies that the HWG eigenproblem must be handled 
with care. This issue is addressed in the HFODD code (see~\cite{(Dob09d)} for further details) by solving the eigenproblem in the 
so-called {\it collective subspace\/} spanned by the {\it natural states\/}:
\begin{equation}   \label{nat_st}   |IM; T_z\rangle^{(m)} =
  \frac{1}{\sqrt{n_m}} \sum_{ij} \eta_{ij}^{(m)}
  {\ket{\varphi_j ; I M; T_z}^{(i)}}.
  \end{equation}
These {\it natural states} are constructed from the eigenstates 
of the norm matrix:
\be\label{egn:eignorm}
\sum_{i'j'} N_{ij ; i'j'} \bar{\eta}^{(m)}_{i'j'} = n_m \bar{\eta}^{(m)}_{ij},
\ee
where, for reasons of numerical stability, only the eigenstates having eigenvalues 
$n= 1,2,...,m_{\text{max}}$ satisfying $n_m > \zeta$ — with $\zeta$ being a 
user-defined basis cut-off parameter — are retained. In the calculations presented below, 
we use typically $\zeta = 0.01$.

The scheme outlined above is used to compute the $|0^+\rangle$ ground states in the mother and daughter nuclei
as well as the intermediate $|1^+_m\rangle$ states and their energies in the $\scn$. 
These are the quantities entering, under the common assumption that Fermi transitions are strongly suppressed in 
the decay process, into the final formula for the nuclear matrix element: 
\begin{widetext}
    \begin{eqnarray}
    \label{nm48}
    \nm\big(\ca \rightarrow \ti \big) = \sum_{m} \frac{\langle \ti; 0^{+}_{f} \vert \sum_{i}\spn_{i}\htapm_{i} \vert \scn; 1_{m}^{+} \rangle \langle \scn; 1_{m}^{+} \vert \sum_{i}\spn_{i}\htapm_{i} \vert \ca; 0^{+}_{i} \rangle}{\Delta E_{m} + \frac{1}{2}Q_{\beta\beta}-\Delta M}. 
\end{eqnarray}
\end{widetext}
Here, the index $m$ runs over the $|1^+_m\rangle$ excitations in $\scn$; $\Delta E_m$ are the calculated excitation energies of the intermediate 
$|1^+_m\rangle$ states in $\scn$ normalized to the lowest experimental $1^+$ state at $\Delta E_1 = 2.2$\,MeV. In the calculation of the matrix element, we use the experimental values forthe $Q$-value  $Q_{\beta\beta} = 4273.7$\,keV and for the mass difference $\Delta M = \Delta M(^{48}\text{Ca}) - \Delta M(^{48}\text{Sc}) = 282$\,keV. The necessary data on mass excess were taken form the NNDC database~\cite{(ensdf_url2)}.

\subsection{Gamow-Teller matrix element within the angular-momentum-projected DFT-NCCI framework}\label{GT-NCCI}

The use of the variant of the DFT-NCCI formalism that involves only  angular-momentum projection introduces certain difficulties 
in the calculation of Gamow-Teller (GT) matrix elements. To illustrate the nature of the problem, let us consider the general GT operator:
\begin{equation}
 \hat {\cal O}^{\rm{(GT)}}_{\mu, \nu} = \hat \tau_{1 \mu} \hat   \sigma_{1\nu}, 
\end{equation}
where $\hat \tau_{1 \mu}$  and $\hat   \sigma_{1\nu}$ are rank-one isospin and spin tensor operators, respectively. 

Within the DFT-NCCI framework, the GT matrix element is expressed as a linear combination of matrix elements evaluated between angular-momentum-projected states:
\begin{widetext}
\begin{eqnarray} \label{M_GT}
  M^{\rm{(GT)}}_{\mu,\nu}  = \langle I' M' K'; T'_z |  
  {\cal O}^{\rm{(GT)}}_{\mu,\nu} | I M K; T_z \rangle & = &
  \langle \varphi (T'_z) | \hat P^{I'}_{K' M'} \hat \tau_{1\mu} \hat   \sigma_{1\nu} \hat P^{I}_{K M} | \psi (T_z) \rangle = \nonumber \\
  & = & 
  C^{I'M'}_{IM, 1\nu} \sum_\xi C^{I'K'}_{IK'-\xi, 1\xi} \langle \varphi | \tau_{1\mu} \hat   \sigma_{1\xi}
   \hat P^{I}_{K'-\xi\, K} | \psi \rangle = \nonumber \\
  & = &  C^{I'M'}_{IM, 1\nu} \sum_\xi C^{I'K'}_{IK'-\xi, 1\xi} \frac{2I+1}{8\pi^2}
   \int d\Omega\, D^{I\, ^*}_{K'-\xi\, K} (\Omega) \, \langle \varphi | \tau_{1\mu} \hat   \sigma_{1\xi}| \widetilde{\psi} \rangle.
\end{eqnarray}
\end{widetext}
Here $ | \varphi (T'_z ) \rangle$ and $ | \psi (T_z ) \rangle$ are Slater determinants corresponding to the daughter and mother nuclei, respectively,  while $ | \widetilde{\psi} \rangle \equiv \hat R(\Omega) | \psi \rangle $ denotes the spatially rotated Slater determinant of the mother nucleus. In the version of the DFT-NCCI approach that includes only angular momentum projection, the configurations $| \varphi (T'_z ) \rangle$ and $| \psi (T_z ) \rangle$ are orthogonal:
\begin{equation}
\langle \varphi (T'_z ) | \psi (T_z) \rangle \sim \delta_{T_z, T'_z} =0 ,
\end{equation}
due to the conservation of neutron and proton numbers. As a consequence, the kernels: 
\begin{equation}\label{kernel}
\langle \varphi | \hat \tau_{1\mu} \hat   \sigma_{1\xi}| \widetilde{\psi} \rangle  
\end{equation}
cannot be directly evaluated, because our implementation of the projection method is based on the Generalized Wick Theorem (GWT), which does not permit the direct calculation of matrix elements between orthogonal Slater determinants.

Nevertheless, this difficulty can be relatively easily overcome. To illustrate the solution, we decompose the Slater determinant into components with good isospin:
\begin{equation}\label{decomp}
| \psi \rangle = \sum_{T\geq T_z} c_T | T T_z\rangle ,
\end{equation}
where
\begin{equation}
| T T_z\rangle = \frac{1}{{\cal N}_T} \hat P^T_{T_z T_z} | \psi \rangle,
\end{equation}
are orthonormal states with the normalization constant
\begin{equation}
{\cal N}_T  =  \langle  \psi |  \hat P^T_{T_z T_z} | \psi \rangle. 
\end{equation}   
The expansion coefficients in Eq.~(\ref{decomp}) are
\begin{equation}
c_T = \langle T T_z | \psi \rangle = \frac{1}{{\cal N}_T}  \langle \psi | \hat P^T_{T_z T_z} | \psi \rangle = \sqrt{{\cal N}_T}
\end{equation}
which allows us to rewrite the Slater determinant $|\psi \rangle$ as follows:
\begin{equation}\label{psi}
|\psi \rangle =  \sum_{T\geq |T_z|}  \hat P^T_{T_z T_z} | \psi \rangle. 
\end{equation} 
Substituting this expression to the kernel~(\ref{kernel}), we obtain:
\begin{widetext}
\begin{eqnarray}
\langle \varphi | \hat \tau_{1\mu} \hat   \sigma_{1\nu} | \widetilde{\psi} \rangle = 
\langle \varphi | \hat \tau_{1\mu} \hat   \sigma_{1\nu} \sum_{T\geq |T_z|}  \hat P^T_{T_z T_z}  | \widetilde{\psi} \rangle =
\sum_{T\geq |T_z|} \frac{2T+1}{2} \int_0^\pi  d\beta_T \sin\beta_T d^T_{T_z T_z} (\beta_T)
\langle \varphi | \hat \tau_{1\mu} \hat   \sigma_{1\nu} \hat R(\beta_T)  | \widetilde{\psi} \rangle .
\end{eqnarray}
\end{widetext}
Here, $\beta_T$ denotes the Euler $\beta$-angle in isospace, and $d^T_{T_z T_z} (\beta_T)$ is the Wigner $d$-function.
In this expression, the Slater determinant on the right is rotated in isospace. This procedure circumvents the issue of orthogonality between Slater determinants representing configurations in two different isobars with different $T_z$ values, at the expense of an additional integration over the $\beta_T$ angle in isospace. The method can be also used to compute Fermi matrix elements. 

To demonstrate the reliability of the method, we have performed several dedicated tests.
For example, by applying it to the $\beta$-decay of the $T_z = -1/2$ nucleus ${}^{21}\text{Na} \rightarrow {}^{21}\text{Ne}$,
we found that only four mesh points are sufficient to obtain a stable solution: $|\mathcal{M}_{\rm{GT}}| = 1.38182$.
This result is very close to the value $|\mathcal{M}_{\rm{GT}}| = 1.37547$, obtained using the full version of the model that includes both angular momentum and isospin projections. This justifies the use of the simplified variant of the model that includes only angular momentum projection. We performed a similar stability tests  for the GT matrix element, 
$\langle {}^{76}\text{As}; 1^+ | \hat {\cal  O}^{\rm{(GT)}}_{-} | {}^{76}\text{Ge}; 0^+\rangle$, 
corresponding to a virtual decay of the $T_z = 6$ nucleus  ${}^{76}\text{Ge}$.  In this 
case, an acceptable approximation $|\mathcal{M}_{\rm{GT}}| = 1.61498$ is obtained already for $n=6$ mesh-points. 
For $n\geq 8$ the calculations yield stable value of $|\mathcal{M}_{\rm{GT}}| = 1.61535$.     


\section{Configuration space of the model}{\label{config}}

The calculations presented here were performed using a developing version of the HFODD 
solver~\cite{(Dob09d),(Sch17),(Dob21)}, which has been equipped with a DFT-NCCI module
 and a subroutine for calculating the GT matrix elements according to the procedure described 
 in Sect.~\ref{GT-NCCI}. In the calculations, we used a spherical basis consisting of 12 harmonic 
 oscillator shells. Integration over the Euler angles $\Omega = (\alpha, \beta, \gamma)$ was 
 performed using Gauss-Chebyshev quadrature (for $\alpha$ and $\gamma$) and Gauss-Legendre 
 quadrature (for $\beta$), with $n_\alpha = n_\beta = n_\gamma = 20$ knots.

The {\it configuration space\/} (and, in turn, the {\it model space\/}) of the DFT-NCCI framework is not fixed, 
as it is in the conventional nuclear shell model. Instead, the configuration space is built step-by-step 
by adding physically relevant low-lying (multi)particle-(multi)hole mean-field configurations, which are 
self-consistent HF solutions conserving parity and signature symmetries. These symmetries are superimposed 
on the mean-field solutions to facilitate convergence of the calculations. This procedure is continued until acceptable 
stability of the calculated observables is achieved.

In our case, we explore configurations built upon single-particle (s.p.) deformed Nilsson levels $|N n_z \Lambda, K ; r\rangle$ 
originating from the spherical $pf$ shell, where $r = \pm i$ is the quantum number associated with the signature-symmetry 
operator $\hat R_y = e^{-i\pi \hat J_y}$. Specifically, we consider proton and neutron Nilsson orbits: 
$|330, 1/2; \pm i\rangle$, $|321, 3/2; \pm i\rangle$, $|312, 5/2; \pm i\rangle$, and $|303, 7/2; \pm i\rangle$ originating from the spherical $f_{7/2}$ subshell; $|310, 1/2; \pm i\rangle$ and $|301, 3/2; \pm i\rangle$ from the spherical $p_{3/2}$ subshell; $|321, 1/2; \pm i\rangle$, $|312, 3/2; \pm i\rangle$, and $|303, 5/2; \pm i\rangle$ from the spherical $f_{5/2}$ subshell; and $|301, 1/2; \pm i\rangle$ from the spherical $p_{1/2}$ subshell.
For reference, in the doubly magic nucleus $^{48}$Ca, the calculated energies of these spherical neutron subshells are: $-11.1$\,MeV ($\nu f_{7/2}$), $-4.0$\,MeV ($\nu p_{3/2}$), $-2.8$\,MeV ($\nu f_{5/2}$), and $-2.1$\,MeV ($\nu p_{1/2}$). 
We performed dedicated tests of the possible influence of the cross $Z(N)=20$ shell excitations on the calculated observables 
which revealed the the effect of these excitatations is negligibly small and they can be safely disregarded.

In the present calculation, we limited ourselves to axial configurations, fixing the orientation of the nucleus with 
its symmetry axis along the $Oy$-axis. This setup allows for a unique association between the single-particle (s.p.) 
level's signature quantum number $r$ and its $K$ quantum number (projection along the symmetry axis) through 
the relation $r = e^{-i\pi K}$. The advantage of having axial symmetry is 
is many-fold. Firstly, it allows for more transparent identification and classification of the configurations. 
Secondly, it allows to use the cranking-around-the-symmetry-axis method which greatly 
facilitates the self-consistent calculation of  the configurations.  Finally, it also allows for simplifying the notation, enabling 
us to label the active Nilsson levels by providing the $K$ quantum number along with an additional subscript to 
differentiate between orbits with the same $K$. Note that the $K$ and $-K$ (denoted below by $\bar{K}$) Nilsson 
levels correspond to opposite signatures. Axial symmetry implies that the total angular-momentum projection onto 
the $Oy$ axis of the intrinsic system, $\Omega = \sum_{i}^{\text{occup}} K_i$, is conserved. Moreover, signature-reversed 
configurations are equivalent.

\subsection{Configurations in $^{48}$Ca}{\label{48Ca-conf}}

The nucleus is doubly magic; hence, the only configurations that can mix with the ground-state configuration are seniority-zero neutron–neutron ($nn$) pairing excitations carrying no alignment, i.e., coupled to $\Omega = 0$. The lowest configurations of this type are $2p$–$2h$ excitations, in which a pair of neutrons is promoted from the $f_{7/2}$ subshell across the $N = 28$ shell gap to the available Nilsson orbitals originating from the $pf_{5/2}$ subshells.

In the model's configuration space, we include all 24 configurations of this type. We disregard the $4p$–$4h$ excitations, except for the lowest one, which is included for testing purposes. The $4p$–$4h$ configurations are expected to play a marginal role, as they are highly excited (well above 20 MeV), mix only weakly with the ground state through the two-body interaction, and are expected to have small Gamow–Teller (GT) matrix elements - since GT is a one-body operator - with the low-lying virtual configurations in $^{48}$Sc.

We also exclude broken-pair, seniority-two $\Omega = 0$ configurations, as they explicitly violate time-reversal symmetry.

\subsection{Configurations in $^{48}$Ti}{\label{48Ti-conf}}

The configuration space of $^{48}$Ti consists of 49 non-aligned ($\Omega = \Omega_\nu + \Omega_\pi = 0$) configurations.
For clarity, it is convenient to divide these configurations into four groups:
\begin{itemize}
\item
    {\bf Group 1} includes the ground-state configuration and the seniority-zero $nn$- and $pp$-pairing $2p$–$2h$ excitations within the $f_{7/2}$ subshell. There are 16 configurations of this type, all of which are included in the calculations.
\item
    {\bf Group 2} consists of 12 $2p$–$2h$ $np$-pairing excitations within the $f_{7/2}$ subshell. All configurations of this type are included.
\item
    {\bf Group 3} contains 12 configurations corresponding to seniority-zero $2p$–$2h$ $nn$-pairing excitations across the $N = 28$ shell gap.
\item
    {\bf Group 4} comprises 9 configurations corresponding to seniority-zero $2p$–$2h$ $pp$-pairing excitations across the $Z = 28$ shell gap.
\end{itemize}

\begin{figure}[h]
\includegraphics[scale = 0.55]{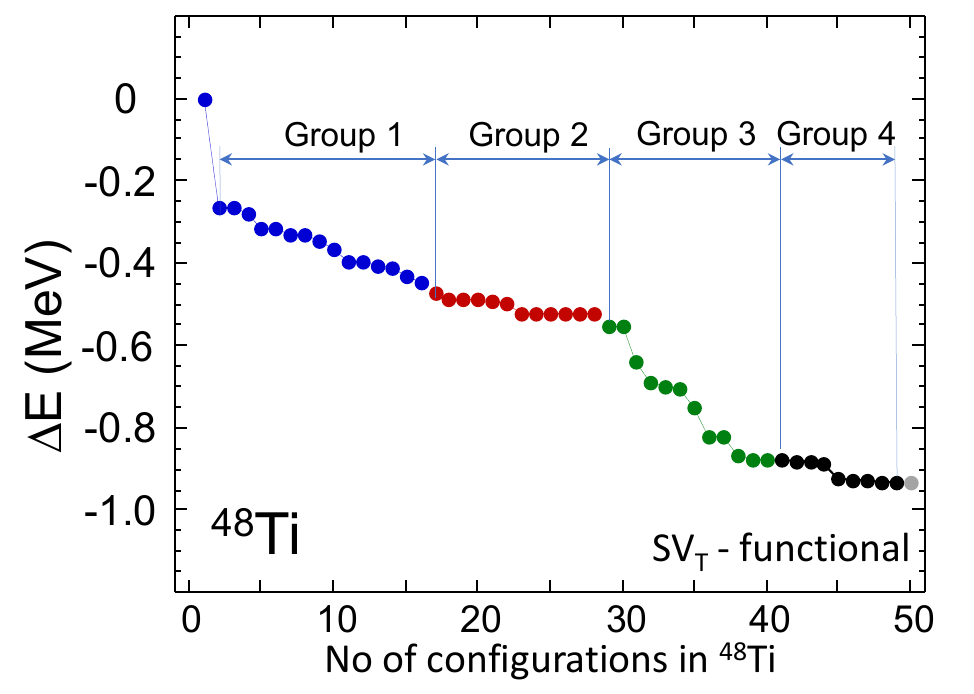}
\caption{\label{48Ti-E} (Color on-line) Correlation energy in the ground state of $^{48}$Ti in function of the number of the 
excited configurations included in the mixing calculation. The configurations are ordered by group 
and, within each group, in ascending order according to their Hartree-Fock energies. The last point represents 
the lowest pp-pairing exciation across the $Z=20$ gap.}
\end{figure}

The calculated energy gain in the ground state of $^{48}$Ti  due to configuration mixing is relatively modest, not exceeding 1\,MeV (see Fig.~\ref{48Ti-E}). As illustrated in the figure, the correlation energy in $^{48}$Ti  saturates reasonably well with the increasing size of the configuration space. The largest contribution to the binding energy comes, as expected, from the configurations belonging to the Groups 1 (all configurations are included) 
and 3, while Group 4 contributes rather marginally.   
A similar energy gain and a similar saturation behavior is calculated for the ground state of $^{48}$Ca.  In spite of the fact that the mixing 
is relatively weak, it appears to play the instrumental role in the calculation of $\nm$ matrix element, as shown below.

\subsection{Configurations in $^{48}$Sc}{\label{48Sc-conf}}

In this case, we assume that the dominant contributions to the Gamow–Teller (GT) matrix element arise from virtual $|1^+\rangle$ states built upon the lowest neutron $1p$–$1h$ and proton  $1p$–$1h$ seniority-two configurations with $|\Omega| = 0, 1$. In total, we include 54 such configurations in the configuration space. These can be grouped into three distinct categories:
\begin{itemize}
\item
    {\bf Group 1} consists of all 11 seniority-two configurations that can be formed within the $f_{7/2}$ shell.
\item
   {\bf  Group 2} includes 26 configurations involving a single neutron excitation across the $N = 28$ shell gap.
\item
    {\bf Group 3} comprises 17 configurations involving a single proton excitation across the $Z = 28$ shell gap.
\end{itemize}

The general strategy is to compute the lowest-energy configurations within each group as the energy denominator 
in Eq.~(\ref{nm48}) damps (albeit quite modestly) the contribution from high-lying $|1^+\rangle $ states to the 
 $\nm$ matrix 
element.

\section{Numerical results}{\label{results}}

\begin{figure}[h]
\vspace{0.2cm}
\centering
\includegraphics[scale = 0.55]{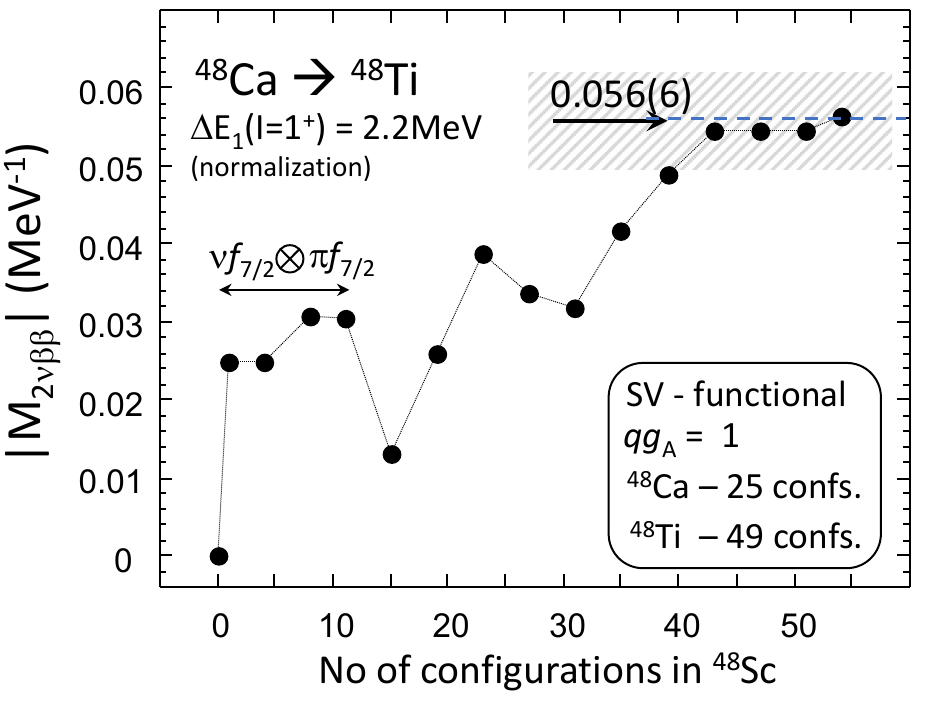}
\caption{\label{vs-48Sc} Stability of the $| \nm |$ matrix element with respect to the number of configurations in 
the virtual nucleus $^{48}$Sc.  The mother and daughter nuclei are fully correlated. The configurations 
in the $^{48}$Sc are in ascending order with respect to their HF excitation energy. The figure shows that the
last (the highest)  14 configurations have marginal impact on the calculated value of the matrix element, much
smaller that the overall theoretical uncertainty marked by the shaded area. 
}
\end{figure}

Fig.~\ref{vs-48Sc} illustrates stability of our calculations against the number of configurations in the virtual nucleus
$^{48}$Sc. In the calculations we fully correlate the mother and daughter nuclei, taking in these nuclei all 
configurations listed in Sect.~\ref{48Ca-conf} and Sect.~\ref{48Ti-conf}, respectively, and gradually increase 
the number of configurations in $^{48}$Sc. The configurations in  $^{48}$Sc are in ascending order. The first 
11 configurations belongs to the Group 1, exhausting all in-shell  $\nu f_{7/2} \otimes \pi f_{7/2}$ seniority-two 
configurations in this nucleus. These strongly linearly dependent configurations account for a bit more than 
50\% of the final value of $\nm \approx 0.056$\, MeV$^{-1}$ of the matrix elements. The remaining contribution 
comes from the cross-shell excitations to the Nilsson orbitals originating from the $pf_{5/2}$ spherical shells.  

\begin{figure}[h]
\vspace{0.2cm}
\centering
\includegraphics[scale=0.55]{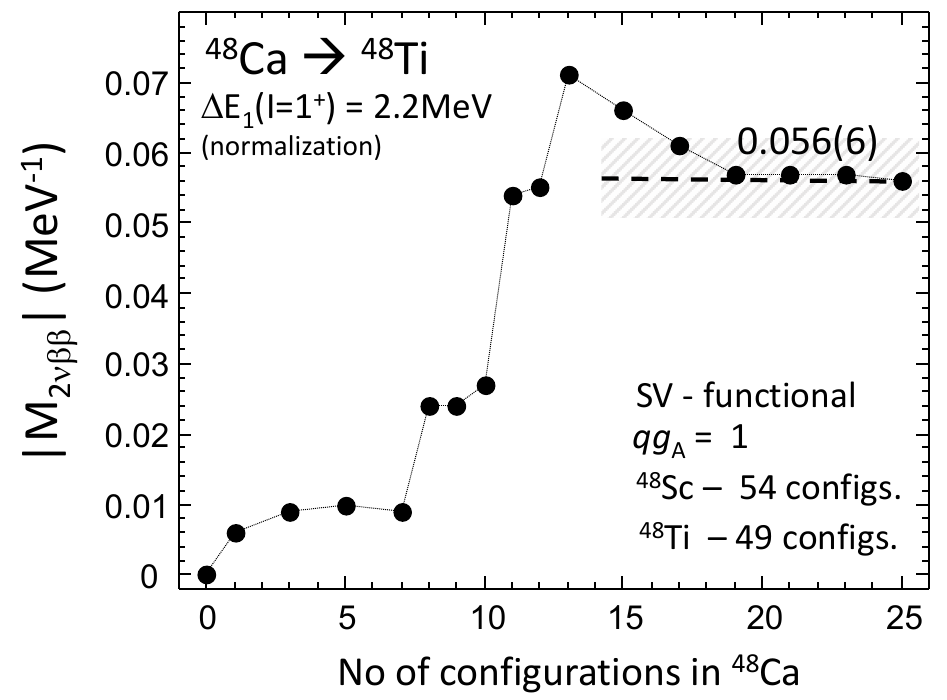}
\caption{\label{vs-48Ca} Stability of the $| \nm |$ matrix element with respect to the number of configurations in 
the mother nucleus $^{48}$Ca.  The virtual and daughter nuclei are fully correlated. The configurations 
in the $^{48}$Ca are in ascending order with respect to their HF excitation energy. The figure shows, unexpectedly,
that the configuration representing the ground state has marginal contribution to the calculated matrix element.  
}
\end{figure}

Fig.~\ref{vs-48Ca} illustrates the stability of the calculations with respect to the number of configurations included in the mother nucleus, $^{48}$Ca, where both the virtual and daughter nuclei are fully correlated. The configurations in $^{48}$Ca are ordered in ascending energy.

It is quite surprising that the contribution from the single configuration representing the mean-field ground state of this doubly magic nucleus is very small in the calculated $|\nm|$. According to our results, the effect arises primarily from small admixtures of neutron $2p$–$2h$ excitations across the $N=28$ magic shell. In contrast, $2p$–$2h$ seniority-zero proton excitations were found to have a negligible impact on the calculated $|\nm|$.
The matrix element reaches stable value of 0.056(6)\, MeV$^{-1}$ already for 18 configurations.  

It is important to emphasize that all possible $2p$–$2h$ seniority-zero neutron excitations were included in this nucleus.

\begin{figure}[h]
\centering
\includegraphics[scale = 0.55]{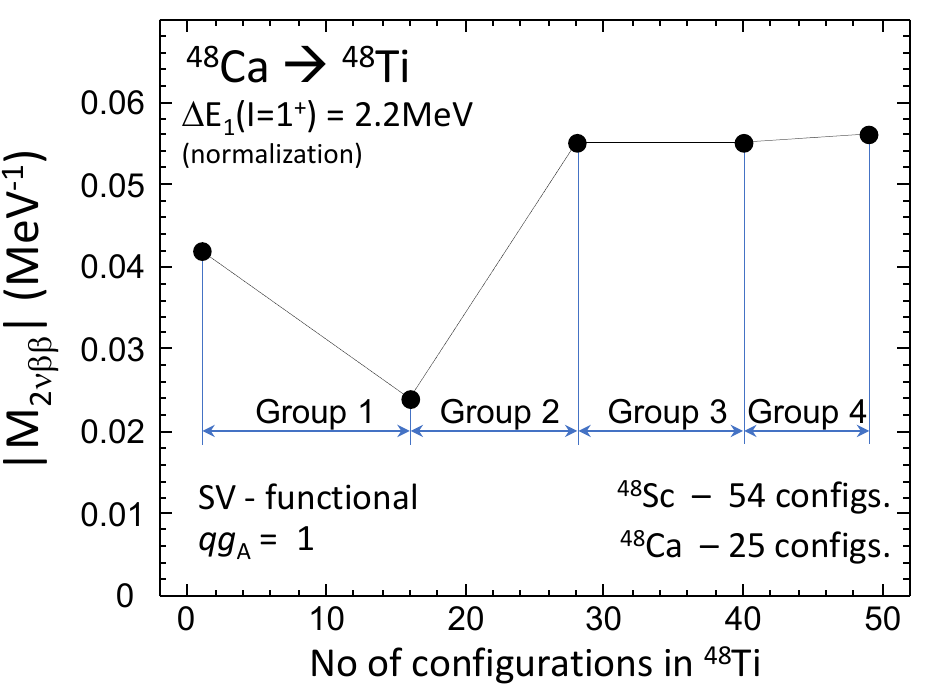}
\caption{\label{vs-48Ti} Stability of the $| \nm |$ matrix element with respect to the number of configurations in 
the daughter nucleus $^{48}$Ti.  The virtual and mother nuclei are fully correlated this time. The configurations 
in the $^{48}$Ti are ordered by group  and, within each group, in ascending order according to their 
Hartree-Fock energies. 
}
\end{figure}

Fig.~\ref{vs-48Ti} illustrates the stability of the calculations with respect to the number of configurations included in the daughter nucleus, $^{48}$Ti. In this case, both the virtual and mother nuclei are fully correlated. The configurations in $^{48}$Ti are ordered differently compared to the two cases discussed above - they are grouped by type. This grouping highlights the impact of different types of configurations on the calculated matrix element. In particular, we observe that in-shell nn/pp pairing correlations - configurations belonging to Group 1 - tend to reduce the value of the matrix element compared to the value obtained using a single configuration representing the ground state of $^{48}$Ti. In contrast, the in-shell np-pairing configurations assigned to Group 2 increase the matrix element, bringing it essentially to its final value. Cross-shell nn/pp-pairing excitations (Groups 3 and 4) affect the value of $| \nm |$ only marginally.

The specific arrangement of configurations used in this study highlights the importance of in-shell np-pairing correlations in determining the final value of the matrix element. This example demonstrates a key advantage of the DFT-NCCI method: built upon the simple and intuitive concepts of the mean field, deformed Nilsson orbitals, and elementary particle-hole configurations, it provides relatively deep insight into the underlying physics - often without the need for a complex analysis of the involved wave functions. 

A natural disadvantage of the method, particularly in comparison to the Nuclear Shell Model (NSM), is that DFT-NCCI, being a global approach applicable to any nucleus on the Segre chart, cannot - and likely never will be - calibrated as precisely as the NSM. In this context, it is therefore especially reassuring to see that our prediction for the nuclear matrix element $\nm$ is in very good agreement with NSM results. Specifically, Ref.~\cite{(Hor07)} reports values of 0.054 (0.064) MeV$^{-1}$ for the GXPF1A (GXPF1) interactions, respectively, while Refs.~\cite{(Iwa15),(Ter21)} yield 0.054 MeV$^{-1}$MeV 
and 0.052 MeV$^{-1}$, respectively. These values, however, are slightly lower than the most recent empirical estimate by Barabash~\cite{(Bar20)}, which, using phase-space factors from Ref.~\cite{(Sto13x)}, yields 0.068(6) MeV$^{-1}$ under the assumption that the effective axial-vector coupling constant is $g_{\text{A}}^{\text{(eff)}} \equiv q g_\text{A} = 1$, where $g_\text{A}\approx 1.27$ denotes the unquenched value.

Our results are subject to theoretical uncertainties, which we estimate to be no greater than 10\%. 
In our opinion, this represents a rather conservative estimate. The calculations appear to be well converged; 
however, as this is a pioneering application of the DFT-NCCI approach to the computation of $\nm$, our experience 
is still limited. Therefore, we cannot rule out the possibility that some relevant configurations are missing from the 
model space, which could affect the results. Another potential source of uncertainty - the normalization of the spectrum 
of $|I=1^+\rangle $ states in the virtual intermediate nucleus - has a moderate impact on the calculated NME. For example, 
changing the normalization from $\Delta E (1^+) = 2.2$\,MeV to 3.0\,MeV changes $|\nm |$ from 0.056\,MeV$^{-1}$
to 0.054\,MeV$^{-1}$.

\begin{figure}[h]
\centering
\includegraphics[scale = 0.55]{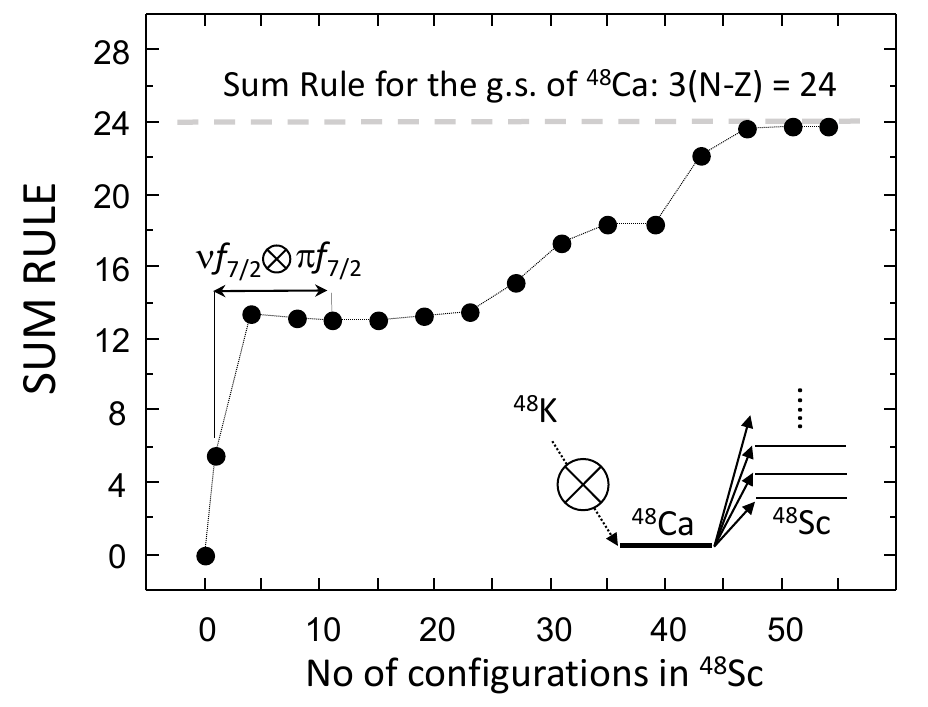}
\caption{\label{Ikeda} (Color on-line) The Ikeda sum rule for the ground-state of the mother nucleus 
 $^{48}$Ca versus a number of configurations taken in the NCCI calculation of this nucleus. 
At each point, i.e. irrespectively on the number of configurations in $^{48}$Ca, we 
include all configurations in the virtual nucleus $^{48}$Sc. The figure neatly underlines the 
role of small admixtures to the dominant mean-field configuration representing ground-state 
of $^{48}$Ca. The insert shows schematically the transitions contributing to the Ikeda sum rule in 
this case.}
\end{figure}

Let us finally comment on the Ikeda sum rules.   Within the DFT-NCCI, computing the Ikeda sum rules requires a dedicated 
calculation, in the analyzed case of the $^{48}$Ca decay in the nuclei $^{48}$K and  $^{48}$V. Fortunately, due to
the structure of $^{48}$K which involves holes in $sd$-shell and particles in $pf_{5/2}$-subshell, the GT decay to
the  $^{48}$Ca is highly forbidden and the sum rule for any state $| \alpha I ; \, T_z \rangle$ in 
$^{48}$Ca is reduced to:
\be
\text{S} = \sum_{\alpha'\, I'}\frac{ | \langle \alpha' I' ; \, T_z - 1 || Y_1^{\text{($-$)}} ||  \alpha I ; \, T_z \rangle |^2}{2I+1} =
3(N-Z) 
\ee 
where
\be
Y^{\text{($\pm$)}} = \sum_i \hat{\vec{\sigma}}^{\text{($i$)}} \hat{t}^{\text{($i$)}}_{\text{($\pm$)}}.  
\ee
The sum rule calculated for the ground state of $^{48}$Ca is illustrated in Fig.~\ref{Ikeda}. 

\section{Summary and conclusions}\label{summary}

\begin{figure}[h]
\centering
\includegraphics[scale = 0.55]{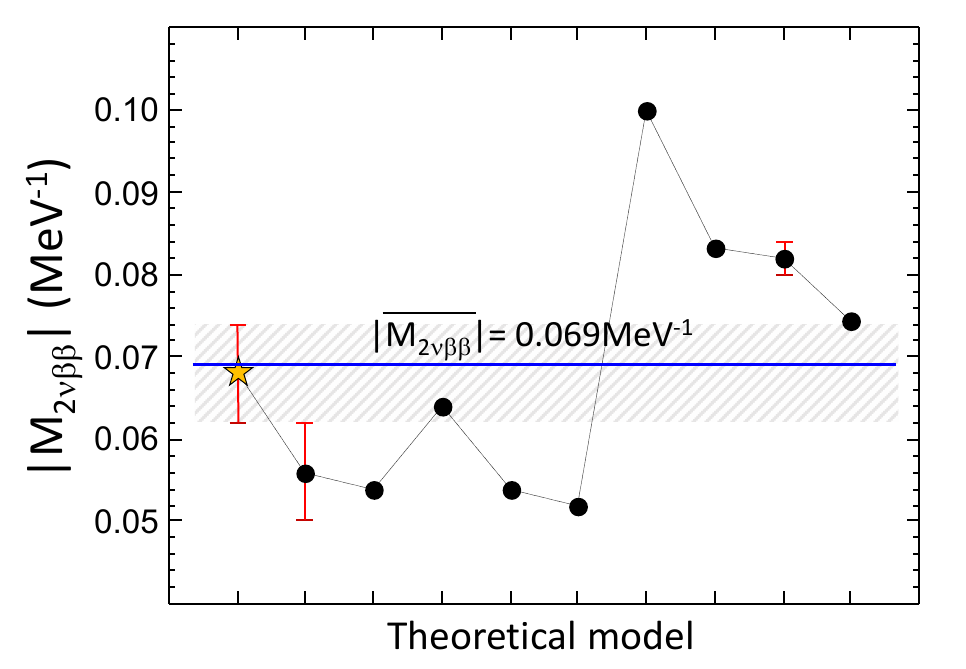}
\caption{\label{MGT-summary} (Color on-line) A summary of selected NME calculated using different models. All values are renormalized to 
the effective axial-vector coupling constant $g^{\text{(eff)}}_{\text{A}} =1$ to emphasize 
the nuclear-structure content. The empirically derived effective NME from 
Ref.~\cite{(Bar20)} is indicated by a star, with its uncertainty shown as a shaded area. The black dots, from left to right, represent: 
the result obtained in this work; two shell-model results from Ref.~\cite{(Hor07)}; shell-model results from Refs.~\cite{(Iwa15)},
~\cite{(Ter21)}, and~\cite{(Kos20)}; and results obtained using alternative (non–shell-model) frameworks reported 
in Refs.~\cite{(Sim18)},~\cite{(Nov21)}, and~\cite{(Ter21)}. See text for further details. The horizontal line indicates the 
average of the theoretical results, which is 0.069$\pm$0.017\,MeV$^{-1}$.
}
\end{figure}

We present a pioneering calculation of the nuclear matrix element for the $2\nu\beta\beta$ decay of ${}^{48}\text{Ca} \rightarrow {}^{48}\text{Ti}$ using DFT-NCCI framework~\cite{(Sat16d)}. In the present calculation, we utilize a variant of the approach that restores rotational symmetry and mixes states projected from self-consistent mean-field configurations obtained by solving the HF equations with the density-independent local 
Skyrme interaction SV$_T$. We propose a novel method of calculating matrix elements of $\beta$-decay operators which detours 
the difficulties associated with the Generalized Wick Theorem (GWT), a framework used in the implementation of projection method 
within the HFODD code, and which does not permit the direct calculation of matrix elements between orthogonal Slater determinants.

Our calculations yield $|\mathcal{M}_{2\nu\beta\beta}| = 0.056(6)$ MeV$^{-1}$ for the NME describing this process. 
The result is in very good agreement with several shell-model studies including 
Ref.~\cite{(Hor07)} which give 0.054 (0.064) MeV$^{-1}$ for the GXPF1A (GXPF1) interactions, 
and  Refs.~\cite{(Iwa15),(Ter21)} which yield 0.054 MeV$^{-1}$MeV  and 0.052 MeV$^{-1}$, respectively.
It also agrees well  with the most recent experimental estimate by Barabash, which is 0.068(6) MeV$^{-1}$, 
assuming quenching $qg_\text{A} \approx 1$.

Other theoretical methods and models, when renormalized to $g_{\text{A}}^{\text{(eff)}} =1$  for comparison purposes, tend to systematically overestimate the empirical matrix element. For example, the shell-model study of Ref.~\cite{(Kos20)} gives 0.100 MeV$^{-1}$, the $pn$QRPA calculation of Ref.~\cite{(Sim18)} yields 0.083 MeV$^{-1}$, the QRPA calculation of Ref.~\cite{(Ter21)} gives 0.075 MeV$^{-1}$, the coupled-cluster calculation 
of Ref.~\cite{(Nov21)} yields 0.082(2) MeV$^{-1}$, and the Second Tamm-Dancoff Approximation (STDA) calculation of Ref.\cite{(Ves25)} gives 0.167 MeV$^{-1}$. These results, with the exception of STDA which is a clear outlier, are shown in Fig.~\ref{MGT-summary}.

Of course, there is no compelling reason to expect all models to use the same effective axial-vector coupling constant, as they operate in different model spaces. Our early DFT-NCCI calculations for mirror $T$=1/2 decays~\cite{(Kon16)} yielded quenching factors consistent with shell-model results in the $sd$- and lower $pf$-shells. However, for the purposes of the present study, more advanced calculations involving larger model spaces are needed. 
To determine a representative $q$-value for the present $\db$ analysis, we plan to compute, among the others, the well measured Gamow-Teller decay of 
the $6^+$ ground state in $^{48}$Sc.

Looking ahead, the DFT-NCCI approach provides a global framework applicable to any nucleus on the Segr\'e chart, allowing for the investigation of NMEs for $\db$  decay in heavier nuclei within a single, unified formalism. At present, we are attempting to compute the NME for the $\db$  decay of 
$^{76}$Ge, which, according to our preliminary calculations, is a triaxial and well-deformed nucleus. 
Our ultimate goal is to generalize and apply the DFT-NCCI framework to the computation of $0\nu\beta\beta$ decay matrix elements. 
Since $\db$ decay constitutes a natural background for the yet unobserved $0\nu\beta\beta$ decay, a deep understanding of $\db$ decay 
in various nuclei is essential for building confidence in the model.

\begin{acknowledgments}

\end{acknowledgments}

\bibliographystyle{apsrev4-2}
\bibliography{apssamp,TEDMED,jacwit34}

\begin{thebibliography}{40}%
\makeatletter
\providecommand \@ifxundefined [1]{%
 \@ifx{#1\undefined}
}%
\providecommand \@ifnum [1]{%
 \ifnum #1\expandafter \@firstoftwo
 \else \expandafter \@secondoftwo
 \fi
}%
\providecommand \@ifx [1]{%
 \ifx #1\expandafter \@firstoftwo
 \else \expandafter \@secondoftwo
 \fi
}%
\providecommand \natexlab [1]{#1}%
\providecommand \enquote  [1]{``#1''}%
\providecommand \bibnamefont  [1]{#1}%
\providecommand \bibfnamefont [1]{#1}%
\providecommand \citenamefont [1]{#1}%
\providecommand \href@noop [0]{\@secondoftwo}%
\providecommand \href [0]{\begingroup \@sanitize@url \@href}%
\providecommand \@href[1]{\@@startlink{#1}\@@href}%
\providecommand \@@href[1]{\endgroup#1\@@endlink}%
\providecommand \@sanitize@url [0]{\catcode `\\12\catcode `\$12\catcode `\&12\catcode `\#12\catcode `\^12\catcode `\_12\catcode `\%12\relax}%
\providecommand \@@startlink[1]{}%
\providecommand \@@endlink[0]{}%
\providecommand \url  [0]{\begingroup\@sanitize@url \@url }%
\providecommand \@url [1]{\endgroup\@href {#1}{\urlprefix }}%
\providecommand \urlprefix  [0]{URL }%
\providecommand \Eprint [0]{\href }%
\providecommand \doibase [0]{https://doi.org/}%
\providecommand \selectlanguage [0]{\@gobble}%
\providecommand \bibinfo  [0]{\@secondoftwo}%
\providecommand \bibfield  [0]{\@secondoftwo}%
\providecommand \translation [1]{[#1]}%
\providecommand \BibitemOpen [0]{}%
\providecommand \bibitemStop [0]{}%
\providecommand \bibitemNoStop [0]{.\EOS\space}%
\providecommand \EOS [0]{\spacefactor3000\relax}%
\providecommand \BibitemShut  [1]{\csname bibitem#1\endcsname}%
\let\auto@bib@innerbib\@empty
\bibitem [{\citenamefont {Avignone}\ \emph {et~al.}(2008)\citenamefont {Avignone}, \citenamefont {Elliott},\ and\ \citenamefont {Engel}}]{(Avi08)}%
  \BibitemOpen
  \bibfield  {author} {\bibinfo {author} {\bibfnamefont {F.~T.}\ \bibnamefont {Avignone}}, \bibinfo {author} {\bibfnamefont {S.~R.}\ \bibnamefont {Elliott}},\ and\ \bibinfo {author} {\bibfnamefont {J.}~\bibnamefont {Engel}},\ }\bibfield  {title} {\bibinfo {title} {{Double} beta decay, {Majorana} neutrinos, and neutrino mass},\ }\href@noop {} {\bibfield  {journal} {\bibinfo  {journal} {Rev. Mod. Phys.}\ }\textbf {\bibinfo {volume} {80}},\ \bibinfo {pages} {481} (\bibinfo {year} {2008})}\BibitemShut {NoStop}%
\bibitem [{\citenamefont {Suhonen}\ and\ \citenamefont {Civitarese}(1998)}]{(Suh98)}%
  \BibitemOpen
  \bibfield  {author} {\bibinfo {author} {\bibfnamefont {J.}~\bibnamefont {Suhonen}}\ and\ \bibinfo {author} {\bibfnamefont {O.}~\bibnamefont {Civitarese}},\ }\bibfield  {title} {\bibinfo {title} {Weak-interaction and nuclear-structure aspects of nuclear double beta decay},\ }\href {https://doi.org/10.1016/S0370-1573(98)00070-1} {\bibfield  {journal} {\bibinfo  {journal} {Physics Reports}\ }\textbf {\bibinfo {volume} {300}},\ \bibinfo {pages} {123} (\bibinfo {year} {1998})}\BibitemShut {NoStop}%
\bibitem [{\citenamefont {Doi}\ \emph {et~al.}(1985)\citenamefont {Doi}, \citenamefont {Kotani},\ and\ \citenamefont {Takasugi}}]{(Doi85)}%
  \BibitemOpen
  \bibfield  {author} {\bibinfo {author} {\bibfnamefont {M.}~\bibnamefont {Doi}}, \bibinfo {author} {\bibfnamefont {T.}~\bibnamefont {Kotani}},\ and\ \bibinfo {author} {\bibfnamefont {E.}~\bibnamefont {Takasugi}},\ }\bibfield  {title} {\bibinfo {title} {Double beta decay and majorana neutrino},\ }\href {https://doi.org/10.1143/PTPS.83.1} {\bibfield  {journal} {\bibinfo  {journal} {Progress of Theoretical Physics Supplement}\ }\textbf {\bibinfo {volume} {83}},\ \bibinfo {pages} {1} (\bibinfo {year} {1985})}\BibitemShut {NoStop}%
\bibitem [{\citenamefont {Tomoda}(1991)}]{(Tom91)}%
  \BibitemOpen
  \bibfield  {author} {\bibinfo {author} {\bibfnamefont {T.}~\bibnamefont {Tomoda}},\ }\bibfield  {title} {\bibinfo {title} {Double beta decay},\ }\href {https://doi.org/10.1088/0034-4885/54/1/002} {\bibfield  {journal} {\bibinfo  {journal} {Reports on Progress in Physics}\ }\textbf {\bibinfo {volume} {54}},\ \bibinfo {pages} {53} (\bibinfo {year} {1991})}\BibitemShut {NoStop}%
\bibitem [{\citenamefont {Konopiski}(1966)}]{(Kon66)}%
  \BibitemOpen
  \bibfield  {author} {\bibinfo {author} {\bibfnamefont {E.}~\bibnamefont {Konopiski}},\ }\href@noop {} {\emph {\bibinfo {title} {The Theory of Beta Radioactivity}}}\ (\bibinfo  {publisher} {Clarendon Press},\ \bibinfo {address} {Oxford},\ \bibinfo {year} {1966})\BibitemShut {NoStop}%
\bibitem [{\citenamefont {Severijns}\ \emph {et~al.}(2006)\citenamefont {Severijns}, \citenamefont {Beck},\ and\ \citenamefont {Naviliat-Cuncic}}]{(Sev06)}%
  \BibitemOpen
  \bibfield  {author} {\bibinfo {author} {\bibfnamefont {N.}~\bibnamefont {Severijns}}, \bibinfo {author} {\bibfnamefont {M.}~\bibnamefont {Beck}},\ and\ \bibinfo {author} {\bibfnamefont {O.}~\bibnamefont {Naviliat-Cuncic}},\ }\bibfield  {title} {\bibinfo {title} {Tests of the standard electroweak model in nuclear beta decay},\ }\href {https://doi.org/10.1103/RevModPhys.78.991} {\bibfield  {journal} {\bibinfo  {journal} {Rev. Mod. Phys.}\ }\textbf {\bibinfo {volume} {78}},\ \bibinfo {pages} {991} (\bibinfo {year} {2006})}\BibitemShut {NoStop}%
\bibitem [{\citenamefont {Erler}\ and\ \citenamefont {Freitas}(2022)}]{(Erl22)}%
  \BibitemOpen
  \bibfield  {author} {\bibinfo {author} {\bibfnamefont {J.}~\bibnamefont {Erler}}\ and\ \bibinfo {author} {\bibfnamefont {A.}~\bibnamefont {Freitas}},\ }\href {https://pdg.lbl.gov/2023/reviews/rpp2022-rev-standard-model.pdf} {\bibinfo {title} {Electroweak model and constraints on new physics}},\ \bibinfo {howpublished} {Online PDF (Particle Data Group)} (\bibinfo {year} {2022}),\ \bibinfo {note} {accessed via PDG: rpp2022 review standard-model}\BibitemShut {NoStop}%
\bibitem [{\citenamefont {Suhonen}(2006)}]{(Suh06)}%
  \BibitemOpen
  \bibfield  {author} {\bibinfo {author} {\bibfnamefont {J.}~\bibnamefont {Suhonen}},\ }\href {https://doi.org/10.1007/978-3-540-48861-3} {\emph {\bibinfo {title} {From Nucleons to Nucleus: Concepts of Microscopic Nuclear Theory}}},\ \bibinfo {edition} {1st}\ ed.,\ Theoretical and Mathematical Physics\ (\bibinfo  {publisher} {Springer Science \& Business Media},\ \bibinfo {address} {Berlin, Heidelberg},\ \bibinfo {year} {2006})\BibitemShut {NoStop}%
\bibitem [{\citenamefont {Herczeg}(2001)}]{(Her01)}%
  \BibitemOpen
  \bibfield  {author} {\bibinfo {author} {\bibfnamefont {P.}~\bibnamefont {Herczeg}},\ }\bibfield  {title} {\bibinfo {title} {Beta decay beyond the standard model},\ }\href {https://doi.org/10.1016/S0146-6410(01)00134-2} {\bibfield  {journal} {\bibinfo  {journal} {Progress in Particle and Nuclear Physics}\ }\textbf {\bibinfo {volume} {46}},\ \bibinfo {pages} {413} (\bibinfo {year} {2001})}\BibitemShut {NoStop}%
\bibitem [{\citenamefont {Aunola}\ \emph {et~al.}(1999)\citenamefont {Aunola}, \citenamefont {Suhonen},\ and\ \citenamefont {Siiskonen}}]{(Aun99)}%
  \BibitemOpen
  \bibfield  {author} {\bibinfo {author} {\bibfnamefont {M.}~\bibnamefont {Aunola}}, \bibinfo {author} {\bibfnamefont {J.}~\bibnamefont {Suhonen}},\ and\ \bibinfo {author} {\bibfnamefont {T.}~\bibnamefont {Siiskonen}},\ }\bibfield  {title} {\bibinfo {title} {Shell-model study of the highly forbidden beta decay $^{48}$ca$\rightarrow ^{48}$sc},\ }\href@noop {} {\bibfield  {journal} {\bibinfo  {journal} {Europhys. Lett.}\ }\textbf {\bibinfo {volume} {46}} (\bibinfo {year} {1999})}\BibitemShut {NoStop}%
\bibitem [{\citenamefont {Boehm}\ and\ \citenamefont {Vogel}(1992)}]{(Vog92)}%
  \BibitemOpen
  \bibfield  {author} {\bibinfo {author} {\bibfnamefont {F.}~\bibnamefont {Boehm}}\ and\ \bibinfo {author} {\bibfnamefont {P.}~\bibnamefont {Vogel}},\ }\href@noop {} {\emph {\bibinfo {title} {Physics of Massive Neutrinos}}}\ (\bibinfo  {publisher} {Cambridge University Press},\ \bibinfo {address} {Cambridge},\ \bibinfo {year} {1992})\BibitemShut {NoStop}%
\bibitem [{\citenamefont {Kondev}\ \emph {et~al.}(2021)\citenamefont {Kondev}, \citenamefont {Wang}, \citenamefont {Huang}, \citenamefont {Naimi},\ and\ \citenamefont {Audi}}]{(Kon21a)}%
  \BibitemOpen
  \bibfield  {author} {\bibinfo {author} {\bibfnamefont {F.~G.}\ \bibnamefont {Kondev}}, \bibinfo {author} {\bibfnamefont {M.}~\bibnamefont {Wang}}, \bibinfo {author} {\bibfnamefont {W.~J.}\ \bibnamefont {Huang}}, \bibinfo {author} {\bibfnamefont {S.}~\bibnamefont {Naimi}},\ and\ \bibinfo {author} {\bibfnamefont {G.}~\bibnamefont {Audi}},\ }\bibfield  {title} {\bibinfo {title} {Beta decay beyond the standard model},\ }\href {https://doi.org/10.1088/1674-1137/abf1d3} {\bibfield  {journal} {\bibinfo  {journal} {Chinese Physics C}\ }\textbf {\bibinfo {volume} {45}},\ \bibinfo {pages} {030001} (\bibinfo {year} {2021})}\BibitemShut {NoStop}%
\bibitem [{\citenamefont {Kotila}\ and\ \citenamefont {Iachello}(2012)}]{(Kot12)}%
  \BibitemOpen
  \bibfield  {author} {\bibinfo {author} {\bibfnamefont {J.}~\bibnamefont {Kotila}}\ and\ \bibinfo {author} {\bibfnamefont {F.}~\bibnamefont {Iachello}},\ }\bibfield  {title} {\bibinfo {title} {Phasespace factors for double$\beta$ decay},\ }\href {https://doi.org/10.1103/PhysRevC.85.034316} {\bibfield  {journal} {\bibinfo  {journal} {Phys. Rev. C}\ }\textbf {\bibinfo {volume} {85}},\ \bibinfo {pages} {034316} (\bibinfo {year} {2012})}\BibitemShut {NoStop}%
\bibitem [{\citenamefont {Stoica}\ and\ \citenamefont {Mirea}(2013)}]{(Sto13x)}%
  \BibitemOpen
  \bibfield  {author} {\bibinfo {author} {\bibfnamefont {S.}~\bibnamefont {Stoica}}\ and\ \bibinfo {author} {\bibfnamefont {M.}~\bibnamefont {Mirea}},\ }\bibfield  {title} {\bibinfo {title} {New calculations for phase space factors involved in double-$\beta$ decay},\ }\href@noop {} {\bibfield  {journal} {\bibinfo  {journal} {Phys. Rev. C}\ }\textbf {\bibinfo {volume} {88}} (\bibinfo {year} {2013})}\BibitemShut {NoStop}%
\bibitem [{\citenamefont {Satu\l{a}}\ \emph {et~al.}(2016)\citenamefont {Satu\l{a}}, \citenamefont {Ba\ifmmode~\mbox{\c{}}\else \c{}\fi{}czyk}, \citenamefont {Dobaczewski},\ and\ \citenamefont {Konieczka}}]{(Sat16d)}%
  \BibitemOpen
  \bibfield  {author} {\bibinfo {author} {\bibfnamefont {W.}~\bibnamefont {Satu\l{a}}}, \bibinfo {author} {\bibfnamefont {P.}~\bibnamefont {Ba\ifmmode~\mbox{\c{}}\else \c{}\fi{}czyk}}, \bibinfo {author} {\bibfnamefont {J.}~\bibnamefont {Dobaczewski}},\ and\ \bibinfo {author} {\bibfnamefont {M.}~\bibnamefont {Konieczka}},\ }\bibfield  {title} {\bibinfo {title} {No-core configuration-interaction model for the isospin- and angular-momentum-projected states},\ }\href@noop {} {\bibfield  {journal} {\bibinfo  {journal} {Phys. Rev. C}\ }\textbf {\bibinfo {volume} {94}},\ \bibinfo {pages} {024306} (\bibinfo {year} {2016})}\BibitemShut {NoStop}%
\bibitem [{\citenamefont {Beiner}\ \emph {et~al.}(1975)\citenamefont {Beiner}, \citenamefont {Flocard}, \citenamefont {Van~Giai},\ and\ \citenamefont {Quentin}}]{(Bei75)}%
  \BibitemOpen
  \bibfield  {author} {\bibinfo {author} {\bibfnamefont {M.}~\bibnamefont {Beiner}}, \bibinfo {author} {\bibfnamefont {H.}~\bibnamefont {Flocard}}, \bibinfo {author} {\bibfnamefont {N.}~\bibnamefont {Van~Giai}},\ and\ \bibinfo {author} {\bibfnamefont {P.}~\bibnamefont {Quentin}},\ }\bibfield  {title} {\bibinfo {title} {{Nuclear} ground-state properties and self-consistent calculations with the skyrme {interaction:(I).} {Spherical} description},\ }\href@noop {} {\bibfield  {journal} {\bibinfo  {journal} {Nucl. Phys. A}\ }\textbf {\bibinfo {volume} {238}},\ \bibinfo {pages} {29} (\bibinfo {year} {1975})}\BibitemShut {NoStop}%
\bibitem [{\citenamefont {Dobaczewski}\ \emph {et~al.}(2007)\citenamefont {Dobaczewski}, \citenamefont {Stoitsov}, \citenamefont {Nazarewicz},\ and\ \citenamefont {Reinhard}}]{(Dob07)}%
  \BibitemOpen
  \bibfield  {author} {\bibinfo {author} {\bibfnamefont {J.}~\bibnamefont {Dobaczewski}}, \bibinfo {author} {\bibfnamefont {M.~V.}\ \bibnamefont {Stoitsov}}, \bibinfo {author} {\bibfnamefont {W.}~\bibnamefont {Nazarewicz}},\ and\ \bibinfo {author} {\bibfnamefont {P.-G.}\ \bibnamefont {Reinhard}},\ }\bibfield  {title} {\bibinfo {title} {Particle-number projection and the density functional theory},\ }\href {https://doi.org/10.1103/PhysRevC.76.054315} {\bibfield  {journal} {\bibinfo  {journal} {Phys. Rev. C}\ }\textbf {\bibinfo {volume} {76}},\ \bibinfo {pages} {054315} (\bibinfo {year} {2007})}\BibitemShut {NoStop}%
\bibitem [{\citenamefont {Satu\l{}a}\ \emph {et~al.}(2009)\citenamefont {Satu\l{}a}, \citenamefont {Dobaczewski}, \citenamefont {Nazarewicz},\ and\ \citenamefont {Rafalski}}]{(Sat09)}%
  \BibitemOpen
  \bibfield  {author} {\bibinfo {author} {\bibfnamefont {W.}~\bibnamefont {Satu\l{}a}}, \bibinfo {author} {\bibfnamefont {J.}~\bibnamefont {Dobaczewski}}, \bibinfo {author} {\bibfnamefont {W.}~\bibnamefont {Nazarewicz}},\ and\ \bibinfo {author} {\bibfnamefont {M.}~\bibnamefont {Rafalski}},\ }\bibfield  {title} {\bibinfo {title} {Isospin mixing in nuclei within the nuclear density functional theory},\ }\href {https://doi.org/10.1103/PhysRevLett.103.012502} {\bibfield  {journal} {\bibinfo  {journal} {Phys. Rev. Lett.}\ }\textbf {\bibinfo {volume} {103}},\ \bibinfo {pages} {012502} (\bibinfo {year} {2009})}\BibitemShut {NoStop}%
\bibitem [{\citenamefont {Satu\l{}a}\ \emph {et~al.}(2011)\citenamefont {Satu\l{}a}, \citenamefont {Dobaczewski}, \citenamefont {Nazarewicz},\ and\ \citenamefont {Rafalski}}]{(Sat11)}%
  \BibitemOpen
  \bibfield  {author} {\bibinfo {author} {\bibfnamefont {W.}~\bibnamefont {Satu\l{}a}}, \bibinfo {author} {\bibfnamefont {J.}~\bibnamefont {Dobaczewski}}, \bibinfo {author} {\bibfnamefont {W.}~\bibnamefont {Nazarewicz}},\ and\ \bibinfo {author} {\bibfnamefont {M.}~\bibnamefont {Rafalski}},\ }\bibfield  {title} {\bibinfo {title} {Microscopic calculations of isospin-breaking corrections to superallowed beta decay},\ }\href {https://doi.org/10.1103/PhysRevLett.106.132502} {\bibfield  {journal} {\bibinfo  {journal} {Phys. Rev. Lett.}\ }\textbf {\bibinfo {volume} {106}},\ \bibinfo {pages} {132502} (\bibinfo {year} {2011})}\BibitemShut {NoStop}%
\bibitem [{\citenamefont {Konieczka}\ \emph {et~al.}(2022)\citenamefont {Konieczka}, \citenamefont {B\k{a}czyk},\ and\ \citenamefont {Satu\l{}a}}]{(Kon22)}%
  \BibitemOpen
  \bibfield  {author} {\bibinfo {author} {\bibfnamefont {M.}~\bibnamefont {Konieczka}}, \bibinfo {author} {\bibfnamefont {P.}~\bibnamefont {B\k{a}czyk}},\ and\ \bibinfo {author} {\bibfnamefont {W.}~\bibnamefont {Satu\l{}a}},\ }\bibfield  {title} {\bibinfo {title} {Precision calculation of isospin-symmetry-breaking corrections to t=1/2 mirror decays using configuration-interaction framework built upon multireference charge-dependent density functional theory},\ }\href {https://doi.org/10.1103/PhysRevC.105.065505} {\bibfield  {journal} {\bibinfo  {journal} {Phys. Rev. C}\ }\textbf {\bibinfo {volume} {105}},\ \bibinfo {pages} {065505} (\bibinfo {year} {2022})}\BibitemShut {NoStop}%
\bibitem [{\citenamefont {B\k{a}czyk}\ \emph {et~al.}(2018)\citenamefont {B\k{a}czyk}, \citenamefont {Dobaczewski}, \citenamefont {Konieczka}, \citenamefont {Satu\l{}a}, \citenamefont {Nakatsukasa},\ and\ \citenamefont {Sato}}]{(Bac18)}%
  \BibitemOpen
  \bibfield  {author} {\bibinfo {author} {\bibfnamefont {P.}~\bibnamefont {B\k{a}czyk}}, \bibinfo {author} {\bibfnamefont {J.}~\bibnamefont {Dobaczewski}}, \bibinfo {author} {\bibfnamefont {M.}~\bibnamefont {Konieczka}}, \bibinfo {author} {\bibfnamefont {W.}~\bibnamefont {Satu\l{}a}}, \bibinfo {author} {\bibfnamefont {T.}~\bibnamefont {Nakatsukasa}},\ and\ \bibinfo {author} {\bibfnamefont {K.}~\bibnamefont {Sato}},\ }\bibfield  {title} {\bibinfo {title} {Isospin-symmetry breaking in masses of $n\approx z$ nuclei},\ }\href {https://arxiv.org/abs/1701.04628v3} {\bibfield  {journal} {\bibinfo  {journal} {Phys. Lett. B}\ }\textbf {\bibinfo {volume} {778}},\ \bibinfo {pages} {178 } (\bibinfo {year} {2018})}\BibitemShut {NoStop}%
\bibitem [{\citenamefont {B\k{a}czyk}\ \emph {et~al.}(2019)\citenamefont {B\k{a}czyk}, \citenamefont {Satu\l{}a}, \citenamefont {Dobaczewski},\ and\ \citenamefont {Konieczka}}]{(Bac19)}%
  \BibitemOpen
  \bibfield  {author} {\bibinfo {author} {\bibfnamefont {P.}~\bibnamefont {B\k{a}czyk}}, \bibinfo {author} {\bibfnamefont {W.}~\bibnamefont {Satu\l{}a}}, \bibinfo {author} {\bibfnamefont {J.}~\bibnamefont {Dobaczewski}},\ and\ \bibinfo {author} {\bibfnamefont {M.}~\bibnamefont {Konieczka}},\ }\bibfield  {title} {\bibinfo {title} {Isobaric multiplet mass equation within nuclear density functional theory},\ }\href {https://arxiv.org/abs/1801.02506} {\bibfield  {journal} {\bibinfo  {journal} {J. Phys. G: Nucl. Part. Phys.}\ }\textbf {\bibinfo {volume} {46}},\ \bibinfo {pages} {03LT01} (\bibinfo {year} {2019})}\BibitemShut {NoStop}%
\bibitem [{\citenamefont {B\k{a}czyk}\ and\ \citenamefont {Satu\l{}a}(2021)}]{(Bac21)}%
  \BibitemOpen
  \bibfield  {author} {\bibinfo {author} {\bibfnamefont {P.}~\bibnamefont {B\k{a}czyk}}\ and\ \bibinfo {author} {\bibfnamefont {W.}~\bibnamefont {Satu\l{}a}},\ }\bibfield  {title} {\bibinfo {title} {Mirror energy differences in $t=1/2\phantom{\rule{4pt}{0ex}}{f}_{7/2}$-shell nuclei within isospin-dependent density functional theory},\ }\href {https://doi.org/10.1103/PhysRevC.103.054320} {\bibfield  {journal} {\bibinfo  {journal} {Phys. Rev. C}\ }\textbf {\bibinfo {volume} {103}},\ \bibinfo {pages} {054320} (\bibinfo {year} {2021})}\BibitemShut {NoStop}%
\bibitem [{\citenamefont {Llewellyn}\ \emph {et~al.}(2020)\citenamefont {Llewellyn}, \citenamefont {Bentley}, \citenamefont {Wadsworth}, \citenamefont {Dobaczewski}, \citenamefont {Satu\l{}a}, \citenamefont {Iwasaki}, \citenamefont {de~Angelis}, \citenamefont {Ash}, \citenamefont {Bazin}, \citenamefont {Bender}, \citenamefont {Cederwall}, \citenamefont {Crider}, \citenamefont {Doncel}, \citenamefont {Elder}, \citenamefont {Elman}, \citenamefont {Gade}, \citenamefont {Grinder}, \citenamefont {Haylett}, \citenamefont {Jenkins}, \citenamefont {Lee}, \citenamefont {Longfellow}, \citenamefont {Lunderberg}, \citenamefont {Mijatovi\'c}, \citenamefont {Milne}, \citenamefont {Rhodes},\ and\ \citenamefont {Weisshaar}}]{(Lle20)}%
  \BibitemOpen
  \bibfield  {author} {\bibinfo {author} {\bibfnamefont {R.~D.~O.}\ \bibnamefont {Llewellyn}}, \bibinfo {author} {\bibfnamefont {M.~A.}\ \bibnamefont {Bentley}}, \bibinfo {author} {\bibfnamefont {R.}~\bibnamefont {Wadsworth}}, \bibinfo {author} {\bibfnamefont {J.}~\bibnamefont {Dobaczewski}}, \bibinfo {author} {\bibfnamefont {W.}~\bibnamefont {Satu\l{}a}}, \bibinfo {author} {\bibfnamefont {H.}~\bibnamefont {Iwasaki}}, \bibinfo {author} {\bibfnamefont {G.}~\bibnamefont {de~Angelis}}, \bibinfo {author} {\bibfnamefont {J.}~\bibnamefont {Ash}}, \bibinfo {author} {\bibfnamefont {D.}~\bibnamefont {Bazin}}, \bibinfo {author} {\bibfnamefont {P.~C.}\ \bibnamefont {Bender}}, \bibinfo {author} {\bibfnamefont {B.}~\bibnamefont {Cederwall}}, \bibinfo {author} {\bibfnamefont {B.~P.}\ \bibnamefont {Crider}}, \bibinfo {author} {\bibfnamefont {M.}~\bibnamefont {Doncel}}, \bibinfo {author} {\bibfnamefont {R.}~\bibnamefont {Elder}}, \bibinfo {author} {\bibfnamefont {B.}~\bibnamefont {Elman}}, \bibinfo {author} {\bibfnamefont
  {A.}~\bibnamefont {Gade}}, \bibinfo {author} {\bibfnamefont {M.}~\bibnamefont {Grinder}}, \bibinfo {author} {\bibfnamefont {T.}~\bibnamefont {Haylett}}, \bibinfo {author} {\bibfnamefont {D.~G.}\ \bibnamefont {Jenkins}}, \bibinfo {author} {\bibfnamefont {I.~Y.}\ \bibnamefont {Lee}}, \bibinfo {author} {\bibfnamefont {B.}~\bibnamefont {Longfellow}}, \bibinfo {author} {\bibfnamefont {E.}~\bibnamefont {Lunderberg}}, \bibinfo {author} {\bibfnamefont {T.}~\bibnamefont {Mijatovi\'c}}, \bibinfo {author} {\bibfnamefont {S.~A.}\ \bibnamefont {Milne}}, \bibinfo {author} {\bibfnamefont {D.}~\bibnamefont {Rhodes}},\ and\ \bibinfo {author} {\bibfnamefont {D.}~\bibnamefont {Weisshaar}},\ }\bibfield  {title} {\bibinfo {title} {Spectroscopy of the $^{79}$zr/$^{79}$y mirror pair},\ }\href {https://doi.org/10.1016/j.physletb.2020.135873} {\bibfield  {journal} {\bibinfo  {journal} {Phys. Lett. B}\ }\textbf {\bibinfo {volume} {811}},\ \bibinfo {pages} {135873} (\bibinfo {year} {2020})}\BibitemShut {NoStop}%
\bibitem [{\citenamefont {Uthayakumaar}\ \emph {et~al.}(2022)\citenamefont {Uthayakumaar}, \citenamefont {Bentley}, \citenamefont {Simpson}, \citenamefont {Haylett}, \citenamefont {Yajzey}, \citenamefont {Lenzi}, \citenamefont {Satu\l{}a}, \citenamefont {Bazin}, \citenamefont {Belarge}, \citenamefont {Bender}, \citenamefont {Davies}, \citenamefont {Elman}, \citenamefont {Gade}, \citenamefont {Iwasaki}, \citenamefont {Kahl}, \citenamefont {Kobayashi}, \citenamefont {Longfellow}, \citenamefont {Lonsdale}, \citenamefont {Lunderberg}, \citenamefont {Morris}, \citenamefont {Napoli}, \citenamefont {Parry}, \citenamefont {Pereira-Lopez}, \citenamefont {Recchia}, \citenamefont {Tostevin}, \citenamefont {Wadsworth},\ and\ \citenamefont {Weisshaar}}]{(Uth22)}%
  \BibitemOpen
  \bibfield  {author} {\bibinfo {author} {\bibfnamefont {S.}~\bibnamefont {Uthayakumaar}}, \bibinfo {author} {\bibfnamefont {M.~A.}\ \bibnamefont {Bentley}}, \bibinfo {author} {\bibfnamefont {E.~C.}\ \bibnamefont {Simpson}}, \bibinfo {author} {\bibfnamefont {T.}~\bibnamefont {Haylett}}, \bibinfo {author} {\bibfnamefont {R.}~\bibnamefont {Yajzey}}, \bibinfo {author} {\bibfnamefont {S.~M.}\ \bibnamefont {Lenzi}}, \bibinfo {author} {\bibfnamefont {W.}~\bibnamefont {Satu\l{}a}}, \bibinfo {author} {\bibfnamefont {D.}~\bibnamefont {Bazin}}, \bibinfo {author} {\bibfnamefont {J.}~\bibnamefont {Belarge}}, \bibinfo {author} {\bibfnamefont {P.~C.}\ \bibnamefont {Bender}}, \bibinfo {author} {\bibfnamefont {P.}~\bibnamefont {Davies}}, \bibinfo {author} {\bibfnamefont {B.}~\bibnamefont {Elman}}, \bibinfo {author} {\bibfnamefont {A.}~\bibnamefont {Gade}}, \bibinfo {author} {\bibfnamefont {H.}~\bibnamefont {Iwasaki}}, \bibinfo {author} {\bibfnamefont {D.}~\bibnamefont {Kahl}}, \bibinfo {author} {\bibfnamefont
  {N.}~\bibnamefont {Kobayashi}}, \bibinfo {author} {\bibfnamefont {B.}~\bibnamefont {Longfellow}}, \bibinfo {author} {\bibfnamefont {S.~J.}\ \bibnamefont {Lonsdale}}, \bibinfo {author} {\bibfnamefont {E.}~\bibnamefont {Lunderberg}}, \bibinfo {author} {\bibfnamefont {L.}~\bibnamefont {Morris}}, \bibinfo {author} {\bibfnamefont {D.~R.}\ \bibnamefont {Napoli}}, \bibinfo {author} {\bibfnamefont {T.}~\bibnamefont {Parry}}, \bibinfo {author} {\bibfnamefont {X.}~\bibnamefont {Pereira-Lopez}}, \bibinfo {author} {\bibfnamefont {F.}~\bibnamefont {Recchia}}, \bibinfo {author} {\bibfnamefont {J.}~\bibnamefont {Tostevin}}, \bibinfo {author} {\bibfnamefont {R.}~\bibnamefont {Wadsworth}},\ and\ \bibinfo {author} {\bibfnamefont {D.}~\bibnamefont {Weisshaar}},\ }\bibfield  {title} {\bibinfo {title} {Spectroscopy of the t=3/2 a=47 and a=45 mirror nuclei via one- and two-nucleon knockout reactions},\ }\href {https://doi.org/10.1103/PhysRevC.106.024327} {\bibfield  {journal} {\bibinfo  {journal} {Phys. Rev. C}\ }\textbf
  {\bibinfo {volume} {106}},\ \bibinfo {pages} {024327} (\bibinfo {year} {2022})}\BibitemShut {NoStop}%
\bibitem [{\citenamefont {Satu\l{}a}\ \emph {et~al.}(2023)\citenamefont {Satu\l{}a}, \citenamefont {Bentley}, \citenamefont {Jalili},\ and\ \citenamefont {Uthayakumaar}}]{(Sat23)}%
  \BibitemOpen
  \bibfield  {author} {\bibinfo {author} {\bibfnamefont {W.}~\bibnamefont {Satu\l{}a}}, \bibinfo {author} {\bibfnamefont {M.~A.}\ \bibnamefont {Bentley}}, \bibinfo {author} {\bibfnamefont {A.}~\bibnamefont {Jalili}},\ and\ \bibinfo {author} {\bibfnamefont {S.}~\bibnamefont {Uthayakumaar}},\ }\bibfield  {title} {\bibinfo {title} {Sensitivity study of mirror energy differences in positive parity bands of $t=3/2$ a=45 nuclei},\ }\href {https://doi.org/10.1103/PhysRevC.108.044315} {\bibfield  {journal} {\bibinfo  {journal} {Phys. Rev. C}\ }\textbf {\bibinfo {volume} {108}},\ \bibinfo {pages} {044315} (\bibinfo {year} {2023})}\BibitemShut {NoStop}%
\bibitem [{\citenamefont {Konieczka}\ \emph {et~al.}(2018)\citenamefont {Konieczka}, \citenamefont {Kortelainen},\ and\ \citenamefont {Satu\l{}a}}]{(Kon18)}%
  \BibitemOpen
  \bibfield  {author} {\bibinfo {author} {\bibfnamefont {M.}~\bibnamefont {Konieczka}}, \bibinfo {author} {\bibfnamefont {M.}~\bibnamefont {Kortelainen}},\ and\ \bibinfo {author} {\bibfnamefont {W.}~\bibnamefont {Satu\l{}a}},\ }\bibfield  {title} {\bibinfo {title} {Gamow-teller response in the configuration space of a density-functional-theory-rooted no-core configuration-interaction model},\ }\href {https://doi.org/10.1103/PhysRevC.97.034310} {\bibfield  {journal} {\bibinfo  {journal} {Phys. Rev. C}\ }\textbf {\bibinfo {volume} {97}},\ \bibinfo {pages} {034310} (\bibinfo {year} {2018})}\BibitemShut {NoStop}%
\bibitem [{\citenamefont {Konieczka}\ \emph {et~al.}(2016)\citenamefont {Konieczka}, \citenamefont {B\k{a}czyk},\ and\ \citenamefont {Satu\l{}a}}]{(Kon16)}%
  \BibitemOpen
  \bibfield  {author} {\bibinfo {author} {\bibfnamefont {M.}~\bibnamefont {Konieczka}}, \bibinfo {author} {\bibfnamefont {P.}~\bibnamefont {B\k{a}czyk}},\ and\ \bibinfo {author} {\bibfnamefont {W.}~\bibnamefont {Satu\l{}a}},\ }\bibfield  {title} {\bibinfo {title} {$\ensuremath{\beta}$-decay study within multireference density functional theory and beyond},\ }\href {https://doi.org/10.1103/PhysRevC.93.042501} {\bibfield  {journal} {\bibinfo  {journal} {Phys. Rev. C}\ }\textbf {\bibinfo {volume} {93}},\ \bibinfo {pages} {042501} (\bibinfo {year} {2016})}\BibitemShut {NoStop}%
\bibitem [{\citenamefont {Dobaczewski}\ \emph {et~al.}(2009)\citenamefont {Dobaczewski}, \citenamefont {Satu{\l}a}, \citenamefont {Carlsson}, \citenamefont {Engel}, \citenamefont {Olbratowski}, \citenamefont {Powa{\l}owski}, \citenamefont {Sadziak}, \citenamefont {Sarich}, \citenamefont {Schunck}, \citenamefont {Staszczak}, \citenamefont {Stoitsov}, \citenamefont {Zalewski},\ and\ \citenamefont {Zdu{\'n}czuk}}]{(Dob09d)}%
  \BibitemOpen
  \bibfield  {author} {\bibinfo {author} {\bibfnamefont {J.}~\bibnamefont {Dobaczewski}}, \bibinfo {author} {\bibfnamefont {W.}~\bibnamefont {Satu{\l}a}}, \bibinfo {author} {\bibfnamefont {B.}~\bibnamefont {Carlsson}}, \bibinfo {author} {\bibfnamefont {J.}~\bibnamefont {Engel}}, \bibinfo {author} {\bibfnamefont {P.}~\bibnamefont {Olbratowski}}, \bibinfo {author} {\bibfnamefont {P.}~\bibnamefont {Powa{\l}owski}}, \bibinfo {author} {\bibfnamefont {M.}~\bibnamefont {Sadziak}}, \bibinfo {author} {\bibfnamefont {J.}~\bibnamefont {Sarich}}, \bibinfo {author} {\bibfnamefont {N.}~\bibnamefont {Schunck}}, \bibinfo {author} {\bibfnamefont {A.}~\bibnamefont {Staszczak}}, \bibinfo {author} {\bibfnamefont {M.}~\bibnamefont {Stoitsov}}, \bibinfo {author} {\bibfnamefont {M.}~\bibnamefont {Zalewski}},\ and\ \bibinfo {author} {\bibfnamefont {H.}~\bibnamefont {Zdu{\'n}czuk}},\ }\href@noop {} {\bibfield  {journal} {\bibinfo  {journal} {Comput. Phys. Commun.}\ }\textbf {\bibinfo {volume} {180}},\ \bibinfo {pages} {2361}
  (\bibinfo {year} {2009})}\BibitemShut {NoStop}%
\bibitem [{(en()}]{(ensdf_url2)}%
  \BibitemOpen
  \href {http://www.nndc.bnl.gov/ensdf/} {}\bibinfo {howpublished} {Evaluated Nuclear Structure Data File, http://www.nndc.bnl.gov/ensdf/}\BibitemShut {NoStop}%
\bibitem [{\citenamefont {Schunck}\ \emph {et~al.}(2017)\citenamefont {Schunck}, \citenamefont {Dobaczewski}, \citenamefont {Satu{\l}a}, \citenamefont {B\k{a}czyk}, \citenamefont {Dudek}, \citenamefont {Gao}, \citenamefont {Konieczka}, \citenamefont {Sato}, \citenamefont {Shi}, \citenamefont {Wang},\ and\ \citenamefont {Werner}}]{(Sch17)}%
  \BibitemOpen
  \bibfield  {author} {\bibinfo {author} {\bibfnamefont {N.}~\bibnamefont {Schunck}}, \bibinfo {author} {\bibfnamefont {J.}~\bibnamefont {Dobaczewski}}, \bibinfo {author} {\bibfnamefont {W.}~\bibnamefont {Satu{\l}a}}, \bibinfo {author} {\bibfnamefont {P.}~\bibnamefont {B\k{a}czyk}}, \bibinfo {author} {\bibfnamefont {J.}~\bibnamefont {Dudek}}, \bibinfo {author} {\bibfnamefont {Y.}~\bibnamefont {Gao}}, \bibinfo {author} {\bibfnamefont {M.}~\bibnamefont {Konieczka}}, \bibinfo {author} {\bibfnamefont {K.}~\bibnamefont {Sato}}, \bibinfo {author} {\bibfnamefont {Y.}~\bibnamefont {Shi}}, \bibinfo {author} {\bibfnamefont {X.}~\bibnamefont {Wang}},\ and\ \bibinfo {author} {\bibfnamefont {T.}~\bibnamefont {Werner}},\ }\bibfield  {title} {\bibinfo {title} {Solution of the skyrme-hartreefockbogolyubovequations in the cartesian deformed harmonic-oscillator basis. (viii) hfodd (v2.73y): A new version of the program},\ }\href {https://doi.org/https://doi.org/10.1016/j.cpc.2017.03.007} {\bibfield  {journal}
  {\bibinfo  {journal} {Computer Physics Communications}\ }\textbf {\bibinfo {volume} {216}},\ \bibinfo {pages} {145 } (\bibinfo {year} {2017})}\BibitemShut {NoStop}%
\bibitem [{\citenamefont {Dobaczewski}\ \emph {et~al.}(2021)\citenamefont {Dobaczewski}, \citenamefont {B\k{a}czyk}, \citenamefont {Becker}, \citenamefont {Bender}, \citenamefont {Bennaceur}, \citenamefont {Bonnard}, \citenamefont {Gao}, \citenamefont {Idini}, \citenamefont {Konieczka}, \citenamefont {Korteleinen}, \citenamefont {Pr\'ochniak}, \citenamefont {Romero}, \citenamefont {Satu{\l}a}, \citenamefont {Shi}, \citenamefont {Werner},\ and\ \citenamefont {Yu}}]{(Dob21)}%
  \BibitemOpen
  \bibfield  {author} {\bibinfo {author} {\bibfnamefont {J.}~\bibnamefont {Dobaczewski}}, \bibinfo {author} {\bibfnamefont {P.}~\bibnamefont {B\k{a}czyk}}, \bibinfo {author} {\bibfnamefont {P.}~\bibnamefont {Becker}}, \bibinfo {author} {\bibfnamefont {M.}~\bibnamefont {Bender}}, \bibinfo {author} {\bibfnamefont {K.}~\bibnamefont {Bennaceur}}, \bibinfo {author} {\bibfnamefont {J.}~\bibnamefont {Bonnard}}, \bibinfo {author} {\bibfnamefont {Y.}~\bibnamefont {Gao}}, \bibinfo {author} {\bibfnamefont {A.}~\bibnamefont {Idini}}, \bibinfo {author} {\bibfnamefont {M.}~\bibnamefont {Konieczka}}, \bibinfo {author} {\bibfnamefont {M.}~\bibnamefont {Korteleinen}}, \bibinfo {author} {\bibfnamefont {L.}~\bibnamefont {Pr\'ochniak}}, \bibinfo {author} {\bibfnamefont {A.}~\bibnamefont {Romero}}, \bibinfo {author} {\bibfnamefont {W.}~\bibnamefont {Satu{\l}a}}, \bibinfo {author} {\bibfnamefont {Y.}~\bibnamefont {Shi}}, \bibinfo {author} {\bibfnamefont {T.}~\bibnamefont {Werner}},\ and\ \bibinfo {author} {\bibfnamefont
  {L.}~\bibnamefont {Yu}},\ }\bibfield  {title} {\bibinfo {title} {Solution of universal nonrelativistic nuclear dft equations in the cartesian deformed harmonic-oscillator basis. (ix) hfodd (v3.06h): a new version of the program},\ }\href {https://doi.org/https://doi.org/10.1016/j.cpc.2017.03.007} {\bibfield  {journal} {\bibinfo  {journal} {Journal of Physics G: Nuclear and Particle Physics}\ }\textbf {\bibinfo {volume} {48}},\ \bibinfo {pages} {102001} (\bibinfo {year} {2021})}\BibitemShut {NoStop}%
\bibitem [{\citenamefont {Horoi}\ \emph {et~al.}(2007)\citenamefont {Horoi}, \citenamefont {Stoica},\ and\ \citenamefont {Brown}}]{(Hor07)}%
  \BibitemOpen
  \bibfield  {author} {\bibinfo {author} {\bibfnamefont {M.}~\bibnamefont {Horoi}}, \bibinfo {author} {\bibfnamefont {S.}~\bibnamefont {Stoica}},\ and\ \bibinfo {author} {\bibfnamefont {B.}~\bibnamefont {Brown}},\ }\bibfield  {title} {\bibinfo {title} {Shell-model calculations of two-neutrino double-$\beta$ decay rates of 48ca with the gxpf1a interaction},\ }\href {https://doi.org/10.1103/PhysRevC.75.034303} {\bibfield  {journal} {\bibinfo  {journal} {Phys. Rev. C}\ }\textbf {\bibinfo {volume} {75}},\ \bibinfo {pages} {034303} (\bibinfo {year} {2007})}\BibitemShut {NoStop}%
\bibitem [{\citenamefont {Iwata}\ \emph {et~al.}(2015)\citenamefont {Iwata}, \citenamefont {Shimizu}, \citenamefont {Utsuno}, \citenamefont {Honma}, \citenamefont {Abe},\ and\ \citenamefont {Otsuka}}]{(Iwa15)}%
  \BibitemOpen
  \bibfield  {author} {\bibinfo {author} {\bibfnamefont {Y.}~\bibnamefont {Iwata}}, \bibinfo {author} {\bibfnamefont {N.}~\bibnamefont {Shimizu}}, \bibinfo {author} {\bibfnamefont {Y.}~\bibnamefont {Utsuno}}, \bibinfo {author} {\bibfnamefont {M.}~\bibnamefont {Honma}}, \bibinfo {author} {\bibfnamefont {T.}~\bibnamefont {Abe}},\ and\ \bibinfo {author} {\bibfnamefont {T.}~\bibnamefont {Otsuka}},\ }\bibfield  {title} {\bibinfo {title} {Ingredients of nuclear matrix element for two-neutrino double-beta decay of 48ca},\ }\href {https://doi.org/10.7566/JPSCP.6.030057} {\bibfield  {journal} {\bibinfo  {journal} {JPS Conf. Proc.}\ }\textbf {\bibinfo {volume} {6}},\ \bibinfo {pages} {030057} (\bibinfo {year} {2015})}\BibitemShut {NoStop}%
\bibitem [{\citenamefont {Terasaki}\ and\ \citenamefont {Iwata}(2021)}]{(Ter21)}%
  \BibitemOpen
  \bibfield  {author} {\bibinfo {author} {\bibfnamefont {J.}~\bibnamefont {Terasaki}}\ and\ \bibinfo {author} {\bibfnamefont {Y.}~\bibnamefont {Iwata}},\ }\bibfield  {title} {\bibinfo {title} {Estimation of nuclear matrix elements of double-$\beta$ decay from shell model and quasiparticle random-phase approximation},\ }\href {https://doi.org/10.1140/epjp/s13360-021-01886-y} {\bibfield  {journal} {\bibinfo  {journal} {Eur. Phys. J. Plus}\ }\textbf {\bibinfo {volume} {136:908}} (\bibinfo {year} {2021})}\BibitemShut {NoStop}%
\bibitem [{\citenamefont {Barabash}(2020)}]{(Bar20)}%
  \BibitemOpen
  \bibfield  {author} {\bibinfo {author} {\bibfnamefont {A.~S.}\ \bibnamefont {Barabash}},\ }\bibfield  {title} {\bibinfo {title} {Precise half-life values for two-neutrino double-$\beta$ decay: 2020 review},\ }\href {https://doi.org/10.3390/universe6100159} {\bibfield  {journal} {\bibinfo  {journal} {Universe}\ }\textbf {\bibinfo {volume} {6}},\ \bibinfo {pages} {159} (\bibinfo {year} {2020})}\BibitemShut {NoStop}%
\bibitem [{\citenamefont {Kostensalo}\ and\ \citenamefont {Suhonen}(2020)}]{(Kos20)}%
  \BibitemOpen
  \bibfield  {author} {\bibinfo {author} {\bibfnamefont {J.}~\bibnamefont {Kostensalo}}\ and\ \bibinfo {author} {\bibfnamefont {J.}~\bibnamefont {Suhonen}},\ }\bibfield  {title} {\bibinfo {title} {Consistent large-scale shell-model analysis of the two-neutrino $\beta\beta$ and single $\beta$ branchings in 48ca and 96zr},\ }\href {https://doi.org/10.1016/j.physletb.2019.135192} {\bibfield  {journal} {\bibinfo  {journal} {Phys. Lett. B}\ }\textbf {\bibinfo {volume} {802}},\ \bibinfo {pages} {135192} (\bibinfo {year} {2020})}\BibitemShut {NoStop}%
\bibitem [{\citenamefont {Simkovic}\ \emph {et~al.}(2018)\citenamefont {Simkovic}, \citenamefont {Dvornicky}, \citenamefont {Stefanik},\ and\ \citenamefont {Faessler}}]{(Sim18)}%
  \BibitemOpen
  \bibfield  {author} {\bibinfo {author} {\bibfnamefont {F.}~\bibnamefont {Simkovic}}, \bibinfo {author} {\bibfnamefont {R.}~\bibnamefont {Dvornicky}}, \bibinfo {author} {\bibfnamefont {D.}~\bibnamefont {Stefanik}},\ and\ \bibinfo {author} {\bibfnamefont {A.}~\bibnamefont {Faessler}},\ }\bibfield  {title} {\bibinfo {title} {Improved description of the $2\nu\beta\beta$-decay and a possibility to determine the effective axial-vector coupling constant},\ }\href {https://doi.org/10.1103/PhysRevC.97.034315} {\bibfield  {journal} {\bibinfo  {journal} {Phys. Rev. C}\ }\textbf {\bibinfo {volume} {97}},\ \bibinfo {pages} {034315} (\bibinfo {year} {2018})}\BibitemShut {NoStop}%
\bibitem [{\citenamefont {Novario}\ \emph {et~al.}(2021)\citenamefont {Novario}, \citenamefont {Gysbergs}, \citenamefont {Engel}, \citenamefont {Hagen}, \citenamefont {Jansen}, \citenamefont {Morris}, \citenamefont {Navratil}, \citenamefont {Papenbrock},\ and\ \citenamefont {Quaglioni}}]{(Nov21)}%
  \BibitemOpen
  \bibfield  {author} {\bibinfo {author} {\bibfnamefont {S.}~\bibnamefont {Novario}}, \bibinfo {author} {\bibfnamefont {P.}~\bibnamefont {Gysbergs}}, \bibinfo {author} {\bibfnamefont {J.}~\bibnamefont {Engel}}, \bibinfo {author} {\bibfnamefont {G.}~\bibnamefont {Hagen}}, \bibinfo {author} {\bibfnamefont {G.}~\bibnamefont {Jansen}}, \bibinfo {author} {\bibfnamefont {T.}~\bibnamefont {Morris}}, \bibinfo {author} {\bibfnamefont {P.}~\bibnamefont {Navratil}}, \bibinfo {author} {\bibfnamefont {T.}~\bibnamefont {Papenbrock}},\ and\ \bibinfo {author} {\bibfnamefont {S.}~\bibnamefont {Quaglioni}},\ }\bibfield  {title} {\bibinfo {title} {Coupled-cluster calculations of neutrinoless double-$\beta$ decay in 48ca},\ }\href {https://doi.org/10.1103/PhysRevLett.126.182502} {\bibfield  {journal} {\bibinfo  {journal} {Phys. Rev. Lett.}\ }\textbf {\bibinfo {volume} {126}},\ \bibinfo {pages} {182502} (\bibinfo {year} {2021})}\BibitemShut {NoStop}%
\bibitem [{\citenamefont {Vesely}\ \emph {et~al.}(2025)\citenamefont {Vesely}, \citenamefont {Denisova}, \citenamefont {Knapp},\ and\ \citenamefont {Simkovic}}]{(Ves25)}%
  \BibitemOpen
  \bibfield  {author} {\bibinfo {author} {\bibfnamefont {P.}~\bibnamefont {Vesely}}, \bibinfo {author} {\bibfnamefont {D.}~\bibnamefont {Denisova}}, \bibinfo {author} {\bibfnamefont {F.}~\bibnamefont {Knapp}},\ and\ \bibinfo {author} {\bibfnamefont {F.}~\bibnamefont {Simkovic}},\ }\bibfield  {title} {\bibinfo {title} {Double-beta decay of 48ca within second tamm--dancoff approximation},\ }\href {https://doi.org/10.5506/APhysPolBSupp.18.2-A9} {\bibfield  {journal} {\bibinfo  {journal} {Acta Phys. Pol. B Proc. Supp.}\ }\textbf {\bibinfo {volume} {18}},\ \bibinfo {pages} {2} (\bibinfo {year} {2025})}\BibitemShut {NoStop}%
\end{thebibliography}%

\end{document}